\pdfoutput=1

\documentclass[useAMS,usenatbib,fleqn]{mnras}

\usepackage{graphicx}
\usepackage{amsmath,amssymb}
\usepackage{natbib}

\DeclareMathOperator{\arcsinh}{arcsinh}

\title[Orbital and escape dynamics in barred galaxies]{Orbital and escape dynamics in barred galaxies - III. \\ The 3D system: Correlations between the basins of escape and the NHIMs}

\author[E. E. Zotos \& Ch. Jung]{Euaggelos E. Zotos$^1$\thanks{E-mail: evzotos@physics.auth.gr} and Christof Jung$^2$\thanks{E-mail: jung@fis.unam.mx} \\
$^1$ Department of Physics, School of Science, Aristotle University of Thessaloniki,
GR-541 24, Thessaloniki, Greece \\
$^2$ Instituto de Ciencias F\'{i}sicas, Universidad Nacional Aut\'{o}noma de M\'{e}xico
Av. Universidad s/n, 62251 Cuernavaca, Mexico
}

\begin{document}

\date{Accepted 2017 September 12. Received 2017 September 12; in original form 2017 August 18}

\pubyear{2018} \volume{473} \pagerange{806–-825}

\setcounter{page}{806}

\maketitle

\label{firstpage}

\begin{abstract}
The escape dynamics of the stars in a barred galaxy composed of a spherically symmetric central nucleus, a bar, a flat thin disk and a dark matter halo component is investigated by using a realistic three degrees of freedom (3-dof) dynamical model. Modern colour-coded diagrams are used for distinguishing between bounded and escaping motion. In addition, the smaller alignment index (SALI) method is deployed for determining the regular, sticky or chaotic nature of bounded orbits. We reveal the basins of escape corresponding to the escape through the two symmetrical escape channels around the Lagrange points $L_2$ and $L_3$ and also we relate them with the corresponding distribution of the escape times of the orbits. Furthermore, we demonstrate how the stable manifolds, around the index-1 saddle points, accurately define the fractal basin boundaries observed in the colour-coded diagrams. The development scenario of the fundamental vertical Lyapunov periodic orbit is thoroughly explored for obtaining a more complete view of the unfolding of the singular behaviour of the dynamics at the cusp values of the parameters. Finally, we examine how the combination of the most important parameters of the bar (such as the semi-major axis and the angular velocity) influences the observed stellar structures (rings and spirals) which are formed by escaping stars guided by the invariant manifolds near the saddle points.
\end{abstract}


\defcitealias{JZ16a}{Paper I}
\defcitealias{JZ16b}{Paper II}

\section{Introduction}
\label{intro}

Usually the important structures of the global dynamics are determined by some unstable invariant subsets of the phase space, where in Hamiltonian systems we can restrict the considerations to a single energy shell. This holds in particular for the escape properties in open systems. Here the escape is basically the process of a motion over a potential saddle which separates an interior potential well from the asymptotic outside region. In \citet{JZ16a} (hereafter \citetalias{JZ16a}) we investigated the escape dynamics of the 2-dof system of a barred galaxy, following similar numerical techniques as in \citet{EP14}.

The most important saddles are the ones at the lowest energy and these are usually index-1 saddles, i.e. saddles where the potential goes down in one direction (one degree of freedom) only. If the energy is close to the saddle energy then usually we find an invariant subset of codimension 2 sitting over this saddle. More explanations and analytic calculations in quadratic approximation for our model system can be found in subsection 4.2 of \citet{JZ16b} (hereafter \citetalias{JZ16b}). These subsets are unstable in the direction in which the potential decreases, this direction is normal to the invariant subset itself and therefore these subsets are generally known under the name ``normally hyperbolic invariant manifolds" (abbreviated NHIMs). The reader can find more general information on NHIMs in \citet{W94}.

For a 3-dof system the codimension 2 NHIMs are 3-dimensional surfaces with the topology of a 3-sphere in the energy shell of the phase space or 2-dimensional surfaces in the domain of the corresponding Poincar\'{e} map. Thereby the restriction of the dynamics to the NHIM itself behaves as a 2-dof dynamics. And then this restricted dynamics is again dominated by some most important lower dimensional subsets. Here, these subsets are periodic orbits in the restricted flow or the corresponding periodic points in the restricted Poincar\'{e} map. For NHIMs over potential saddles these important periodic orbits are the Lyapunov periodic orbits \citep{L07}. As has been seen in \citetalias{JZ16b}, in the special case of our potential model for the barred galaxy the vertical Lyapunov orbit over the index-1 saddles of the effective potential in the rotating coordinate system plays the role of the most important element of the dynamics. In the following this particular periodic orbit will be called $\Gamma_v$. Its behaviour and its development scenario determine, to an amazingly large extent, the global dynamics of the galaxy system.

In \citetalias{JZ16b} the development scenario of $\Gamma_v$ in particular and also of the whole saddle NHIM has been studied as function of the energy $E$, but only for fixed values of the bar length and of the rotation velocity of the bar. Therefore the first pending problem is a detailed study of the scenario of $\Gamma_v$ also under varying values of these bar parameters. A second pending problem has been the clarification of the exact nature of some singularities found in the scenario of $\Gamma_v$. Fortunately the solution of this second problem will be provided automatically by the solution of the first problem. If we embed a singular behaviour into a parameter space of sufficient dimension then the singularity will be unfolded and its nature becomes evident \citep[see e.g.,][]{PS78}. And it will turn out that the embedding of the development scenario of $\Gamma_v$ under a change of $E$ into the space of the two additional bar parameters does this job. In this sense one of the main tasks of the present work will be the presentation of the development scenario of $\Gamma_v$ in the 3-dimensional parameter space.

When we study the dependence of the scenario on the bar parameters then also the question comes up how the rings and spirals formed by the outer branches of the unstable manifolds of the NHIMs depend on these bar parameters. This is important since the rings and spirals in barred galaxies are observational properties. So we obtain connections between the parameters in our model and observational results \citep[e.g.,][]{ARGM09,ARGBM09,ARGBM10,ARGM11,RGMA06,RGAM07}.

Another interesting problem still pending is the following: In the interior potential well we find initial conditions which lead to bound regular motion, we find other initial conditions leading to escape over one or the other potential saddle after moderate time and we find initial conditions for orbits which only escape after an extremely long time after they have performed vast chaotic bound transients or have stuck for a very long time to the fractal surface of islands of bound motion. If according to our claim the saddle NHIMs direct the global escape process, then these various subsets of behaviour and in particular their basin boundaries should be determined by the properties of the saddle NHIMs. General information on the basin boundaries can be found in chapter 5 of \citet{LT11}, while for an instructive example of application to escape in a scattering system see subsection 8.5.4 of the same book. Additional information regarding the basin boundaries in chaotic scattering are given in  section 3 of \citet{SS12}.

The present article is the last part of the series and it is organized as follows: In Section \ref{galmod} we provide an analytic description of the main properties of the barred galaxy model. In the following Section we conduct a thorough numerical investigation of the escape dynamics of stars, while in Section \ref{orbdyn} we explore the orbital dynamics of the 3-dof system. In Section \ref{bbs} we show in which way the basin boundaries are given by the stable manifolds of the codimension-2 saddle NHIMs. In Section \ref{fvo} we present the development scenario of the fundamental vertical orbit $\Gamma_v$. In Section \ref{spr} we link the invariant manifolds with the stellar structures formed by escaping stars, while in Section \ref{disc} the discussion is given.

\section{Presentation of the galactic model}
\label{galmod}

In \citet{JZ15} we introduced a new realistic multi-component model for describing the motion of stars in barred galaxies, while in \citetalias{JZ16a} we upgraded this model by adding a dark matter halo component. The total gravitational potential $\Phi_{\rm t}(x,y,z)$ is composed of the following components:
\begin{itemize}
  \item A spherically symmetric nucleus described by a Plummer potential \citep{BT08}
  \begin{equation}
  \Phi_{\rm n}(x,y,z) = - \frac{G M_{\rm n}}{\sqrt{x^2 + y^2 + z^ 2 + c_{\rm n}^2}},
  \label{Vn}
  \end{equation}
  where $G$ is the gravitational constant, while $M_{\rm n}$ and $c_{\rm n}$ are the mass and the scale length of the nucleus, respectively.
  \item A rotating bar modelled by the analytical potential
  \begin{align}
  \Phi_{\rm b}(x,y,z) &= \frac{G M_{\rm b}}{2a}\left[\arcsinh \left( \frac{x-a}{d} \right) - \arcsinh \left( \frac{x+a}{d} \right) \right] = \nonumber \\
  &= \frac{G M_{\rm b}}{2a} \ln \left( \frac{x-a+\sqrt{(x-a)^2 + d^2}} {x+a+\sqrt{(x+a)^2 + d^2}} \right),
  \label{Vb}
  \end{align}
  where $d^2 = y^2 + z^2 + c_{\rm b}^2$, $M_{\rm b}$ is the mass of the bar, $a$ is the length of the semi-major axis of the bar, while $c_{\rm b}$ is its scale length. In \citet{JZ15} (see Eq. (6)) we have given in closed form the mass density belonging to this new model potential. This equation shows immediately that the corresponding mass density is always and everywhere positive, independently of the values of all parameters and in particular of the length parameter $a$. Accordingly, from this side there are no objections against any variation of all the bar parameters.
  \item A flat thin disc described by a Miyamoto-Nagai potential \citep{MN75}
  \begin{equation}
  \Phi_{\rm d}(x,y,z) = - \frac{G M_{\rm d}}{\sqrt{x^2 + y^2 + \left(k + \sqrt{h^2 + z^ 2}\right)^2}},
  \label{Vd}
  \end{equation}
  where $M_{\rm d}$ is the mass of the disc, while $k$ and $h$ are the horizontal and vertical scale lengths of the disc, respectively.
  \item A spherically symmetric dark matter halo described by a Plummer potential
  \begin{equation}
  \Phi_{\rm h}(x,y,z) = - \frac{G M_{\rm h}}{\sqrt{x^2 + y^2 + z^ 2 + c_{\rm h}^2}},
  \label{Vh}
  \end{equation}
  where $M_{\rm h}$ and $c_{\rm h}$ are the mass and the scale length of the dark matter halo, respectively.
\end{itemize}

It is assumed that the galactic bar rotates clockwise, around the vertical $z$-axis, at a constant angular velocity $\Omega_{\rm b}$. Therefore, the dynamics of the system can be described in a rotating reference frame where the semi-major axis of the bar points into the $x$ direction, while its intermediate axis points into the $y$ direction. On this basis, the corresponding effective potential reads
\begin{equation}
\Phi_{\rm eff}(x,y,z) = \Phi_{\rm t}(x,y,z) - \frac{1}{2}\Omega_{\rm b}^2 \left(x^2 + y^2 \right).
\label{Veff}
\end{equation}

We use the same system of units which was also used in \citetalias{JZ16a}, where
\begin{itemize}
  \item The length unit is 1 kpc.
  \item The mass unit is $2.325 \times 10^7 {\rm M}_\odot$.
  \item The time unit is $0.9778 \times 10^8$ yr (about 100 Myr).
  \item The velocity unit is 10 km $s^{-1}$.
  \item The angular momentum unit (per unit mass) is 10 km kpc s$^{-1}$.
  \item The energy unit (per unit mass) is 100 km$^2$s$^{-2}$.
\end{itemize}
In the above-mentioned units the values of the involved parameters are: $M_{\rm n} = 400$ (corresponding to 9.3 $\times 10^{9}$ ${\rm M}_\odot$), $c_{\rm n} = 0.25$ kpc, $M_{\rm b} = 3500$ (corresponding to 8.13 $\times 10^{10}$ ${\rm M}_\odot$), $a = 10$ kpc, $c_{\rm b} = 1$ kpc, $M_{\rm d} = 7000$ (corresponding to 1.6275 $\times 10^{11}$ ${\rm M}_\odot$), $k = 3$ kpc, $h = 0.175$ kpc, $M_{\rm h} = 20000$ (corresponding to 4.65 $\times 10^{11}$ ${\rm M}_\odot$), $c_{\rm h} = 20$ kpc, and $\Omega_{\rm b} = 4.5$. This set of the values of the parameters defines the so-called Standard Model (SM).

The equations of motion are
\begin{align}
\dot{x} &= p_x + \Omega_{\rm b} y, \nonumber \\
\dot{y} &= p_y - \Omega_{\rm b} x, \nonumber \\
\dot{z} &= p_z, \nonumber \\
\dot{p_x} &= - \frac{\partial \Phi_{\rm t}}{\partial x} + \Omega_{\rm b} p_y, \nonumber \\
\dot{p_y} &= - \frac{\partial \Phi_{\rm t}}{\partial y} - \Omega_{\rm b} p_x, \nonumber \\
\dot{p_z} &= - \frac{\partial \Phi_{\rm t}}{\partial z},
\label{eqmot}
\end{align}
where of course the dot indicates the derivative with respect to the time.

Similarly the variational equations, which govern the evolution of a deviation vector ${\vec{w}} = (\delta x, \delta y, \delta z, \delta p_x, \delta p_y, \delta p_z)$, are
\begin{align}
\dot{(\delta x)} &= \delta p_x + \Omega_{\rm b} \delta y, \nonumber \\
\dot{(\delta y)} &= \delta p_y - \Omega_{\rm b} \delta x, \nonumber \\
\dot{(\delta z)} &= \delta p_z, \nonumber \\
(\dot{\delta p_x}) &=
- \frac{\partial^2 \Phi_{\rm t}}{\partial x^2} \ \delta x
- \frac{\partial^2 \Phi_{\rm t}}{\partial x \partial y} \delta y
- \frac{\partial^2 \Phi_{\rm t}}{\partial x \partial z} \delta z + \Omega_{\rm b} \delta p_y, \nonumber \\
(\dot{\delta p_y}) &=
- \frac{\partial^2 \Phi_{\rm t}}{\partial y \partial x} \delta x
- \frac{\partial^2 \Phi_{\rm t}}{\partial y^2} \delta y
- \frac{\partial^2 \Phi_{\rm t}}{\partial y \partial z} \delta z - \Omega_{\rm b} \delta p_x, \nonumber \\
(\dot{\delta p_z}) &=
- \frac{\partial^2 \Phi_{\rm t}}{\partial z \partial x} \delta x
- \frac{\partial^2 \Phi_{\rm t}}{\partial z \partial y} \delta y
- \frac{\partial^2 \Phi_{\rm t}}{\partial z^2} \delta z.
\label{vareq}
\end{align}

The corresponding Hamiltonian, which determines the three-dimensional motion of a test-particle (star) with a unit mass $(m = 1)$, is
\begin{align}
H(x,y,z,p_x,p_y,p_z) &= \frac{1}{2} \left(p_x^2 + p_y^2 + p_z^2 \right) + \Phi_{\rm t}(x,y,z) \nonumber\\
&- \Omega_{\rm b} L_z = E,
\label{ham}
\end{align}
where $p_x$, $p_y$ and $p_z$ are the canonical momenta per unit mass, conjugate to $x$, $y$ and $z$ respectively, $E$ is the numerical value of the Jacobi integral of motion, which is conserved, while $L_z = x p_y - y p_x$ is the component of the angular momentum along the vertical $z$ direction.

\begin{figure}
\begin{center}
\includegraphics[width=\hsize]{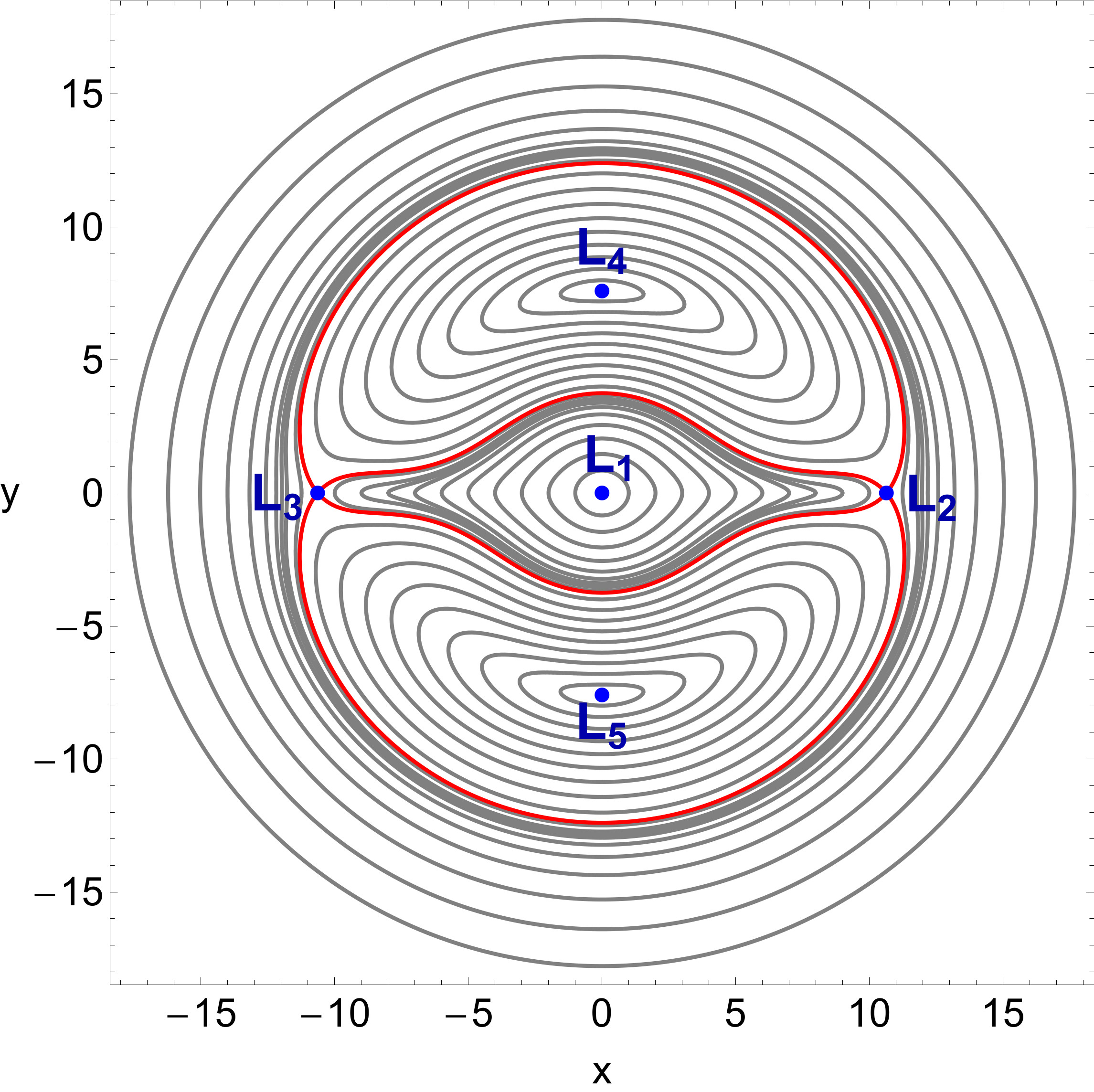}
\end{center}
\caption{The isoline contours of the effective potential on the $(x,y)$ plane (when $z = 0$) for the standard model. The positions of the five Lagrange points are indicated by blue dots. The isoline contours corresponding to the critical energy of escape $E(L_2)$ are shown in red. (For the interpretation of references to colour in this figure caption and the corresponding text, the reader is referred to the electronic version of the article.)}
\label{isoc}
\end{figure}

As we know, for the standard model (standard values of the bar parameters) the non-linear system of the barred galaxy has five equilibrium points at which the first order partial derivatives are zero
\begin{equation}
\frac{\partial \Phi_{\rm eff}}{\partial x} = \frac{\partial \Phi_{\rm eff}}{\partial y} = \frac{\partial \Phi_{\rm eff}}{\partial z} = 0.
\label{lgs}
\end{equation}
These equilibrium points are also know as Lagrange points. The central stationary point $L_1$, located at $(x,y,z) = (0,0,0)$, is a local minimum of the effective potential. The Lagrange points $L_2$ and $L_3$ are index-1 saddle points of the effective potential located at $(x,y,z) = (\pm r_L,0,0)$, where $r_L = 10.63695596$ is the Lagrange radius. On the other hand, the libration points $L_4$ and $L_5$ are index-2 saddle points of the effective potential. Fig. \ref{isoc} shows the isoline contours of constant effective potential on the $(x,y)$ plane (when $z = 0$). The positions of the five equilibrium points are also indicated in the same figure.

The numerical values of the effective potential at the index-1 saddle points $L_2$, $L_3$ as well as at the index-2 saddle points $L_4$ and $L_5$ are critical values of the Hamiltonian. For the standard model we have $E(L_2) = -3242.77217493$ and $E(L_4) = -2800.50348529$ (remember that $E(L_2) = E(L_3)$ and $E(L_4) = E(L_5)$). In fact $E(L_2)$ is the energy of escape of the system. This is true because when $E > E(L_2)$ the zero velocity surfaces open and two symmetrical channels of escape emerge in the vicinity of the the libration points $L_2$ and $L_3$. Through these exits the stars are allowed to enter the exterior realm (when $x < -r_L$ or when $x > +r_L$) and therefore are free to escape to infinity.

A double precision Bulirsch-Stoer algorithm, written in standard \verb!FORTRAN 77! \citep[see e.g.,][]{PTVF92}, was used for the numerical integration of both the sets of the equations of motion (\ref{eqmot}) and the variational equations (\ref{vareq}). The integrator has a fixed time step equal to $10^{-2}$, which is sufficient for the desired accuracy of our calculations. The numerical error, regarding the conservation of the value of the total orbital energy given by Eq. (\ref{ham}), was found, in most cases, to be smaller than $10^{-14}$. The latest version 11.1.1 of Mathematica$^{\circledR}$ \citep[e.g.,][]{Wolf03} was used for preparing all the graphical illustration of the paper.

\section{Escape dynamics}
\label{escdyn}

In \citetalias{JZ16a} we investigated the escape dynamics of the 2-dof system, that is when $z = 0$. Now we reveal the escape dynamics of the full 3-dof system. In particular, it would be very interesting to determine how the initial value of the $z$ coordinate of the orbits influences the overall escape process of the stars. For this task we define, for several values of the total orbital energy $E$, dense uniform grids of $1024 \times 1024$ initial conditions on the $(x,z)$ plane. Always the initial conditions of the orbits lie inside the energetically allowed area (when $\Phi_{\rm eff}(x,y,z) < E$) and also inside the Lagrange radius with $\sqrt{x_0^2 + y_0^2 + z_0^2} \leq r_L$. All orbits have initial conditions  $(x_0,z_0)$ with $y_0 = p_{x0} = p_{z0} = 0$, while $p_{y0}$ is always obtained from the Jacobi integral of motion (\ref{ham}). Here we note that in our computations we choose the branch of $p_y$ with positive orientation of intersection of the plane $y = 0$.

Our investigation will be focused on the energy interval $E \in (E(L_2),E(L_4)]$. We restrict our numerical analysis to this particular energy interval because if the energy is higher than $E(L_4)$ then the saddles, over the Lagrange points $L_2$ and $L_3$, are no longer important for the escape of the stars.

The escape of a test particle (star) can easily be determined by using a simple geometrical escape criterion. In particular, an orbit is considered to escape from the scattering region when the star passes over one of the Lagrange points $L_2$ or $L_3$, with the velocity pointing outwards.

In the following we will distinguish between bounded and escaping (unbounded) motion of stars. In earlier related works \citep[see e.g.,][]{Z15,ZJ17} a substantial amount of trapped chaotic orbits has been detected. These orbits were found to escape from the system only after extremely long time intervals of numerical integration. On this basis, it would be beneficial to distinguish between non-escaping regular orbits and trapped chaotic ones.

Several dynamical indicators exist which can help us to determine the ordered or chaotic character of an orbit. As in all the previous papers of the series, our choice is the Smaller Alignment Index (SALI) method \citep{S01} which is, beyond any doubt, a very fast and reliable tool. The value of SALI at the end of the numerical integration suggests the nature of the orbit. More precisely according \cite{SABV04}, if SALI $> 10^{-4}$ the orbit is regular, while if SALI $< 10^{-8}$ the orbit is surely chaotic. Furthermore, when $10^{-8} \leq \rm SALI \leq 10^{-4}$ the orbit under investigation is a sticky orbit and further numerical integration is required in order to fully reveal its true chaotic character.

A maximum integration time of $10^4$ dimensionless time units is set for all the initial conditions of the orbits. We decided to set such a long integration time so as to be sure that all orbits have enough time to escape, through one of the escape channels. Note, that the same time limit has also been used in \citetalias{JZ16a}. At this point, we would like to clarify that any orbits which remain trapped inside the Lagrange radius after $10^4$ time units are considered as a trapped (ordered or chaotic) orbits, even if in the long run (integrated for $t > 10^4$ time units) they do escape.

The initial conditions of the three-dimensional orbits will be classified into five main categories:
\begin{enumerate}
  \item Non-escaping regular orbits.
  \item Trapped sticky orbits.
  \item Trapped chaotic orbits.
  \item Escaping orbits through $L_2$.
  \item Escaping orbits through $L_3$.
\end{enumerate}

\begin{figure*}
\centering
\resizebox{0.75\hsize}{!}{\includegraphics{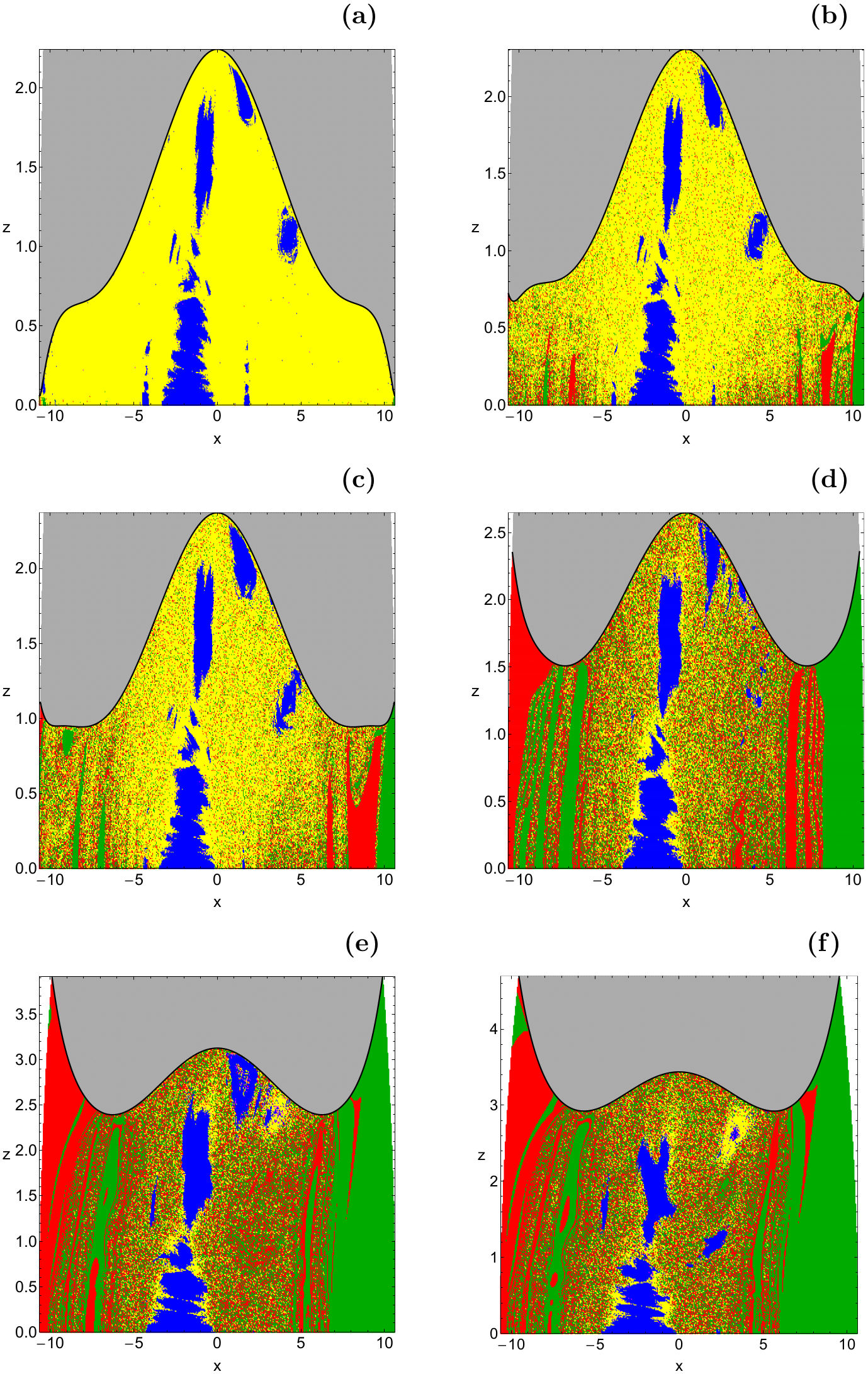}}
\caption{Orbital structure of the $(x,z)$ plane when (a): $E = -3240$; (b): $E = -3211$; (c): $E = -3182$; (d): $E = -3068$; (e): $E = -2895$; (f): $E = E(L_4)$. The colour code is as follows: non-escaping regular orbits (blue), trapped sticky orbits (magenta), trapped chaotic orbits (yellow), escaping orbits through $L_2$ (green), escaping orbits through $L_3$ (red). The energetically forbidden areas are marked with grey. (For the interpretation of references to colour in this figure caption and the corresponding text, the reader is referred to the electronic version of the article.)}
\label{xz}
\end{figure*}

\begin{figure*}
\centering
\resizebox{0.85\hsize}{!}{\includegraphics{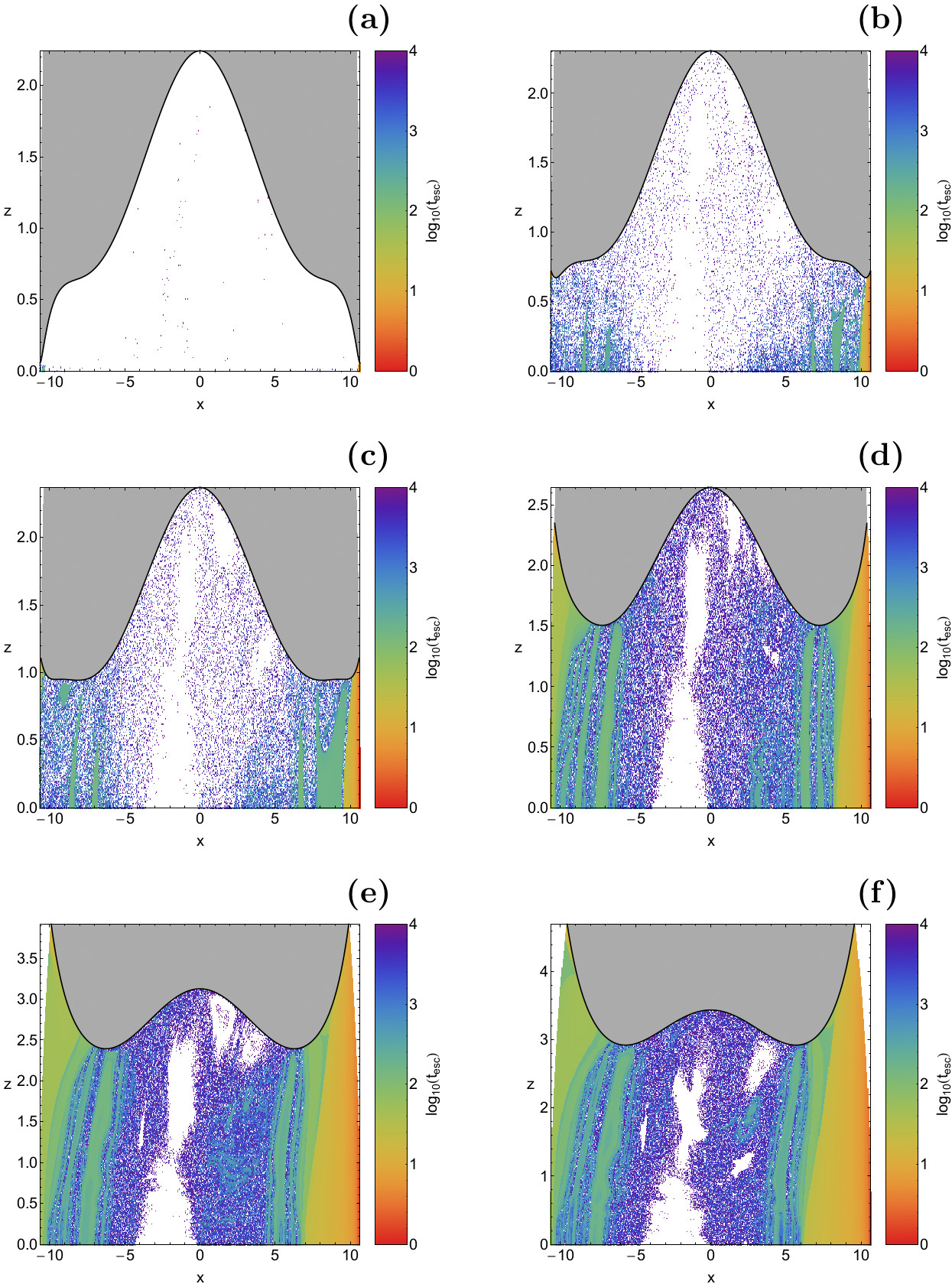}}
\caption{Distribution of the corresponding escape time $t_{\rm esc}$ of the orbits on the $(x,z)$ plane for the values of the total orbital energy presented in panels (a-f) of Fig. \ref{xz}, respectively. The darker the colour, the higher the escape time. Initial conditions of non-escaping regular orbits as well as of trapped sticky and chaotic orbits are shown in white. Note that the colour bar has a logarithmic scale. (For the interpretation of references to colour in this figure caption and the corresponding text, the reader is referred to the electronic version of the article.)}
\label{xzt}
\end{figure*}

In Fig. \ref{xz}(a-f) we present the orbital structure of the $(x,z)$ plane for six values of the total energy $E$. In these colour-coded grids (CCDs) we assign to each pixel a specific colour according to the corresponding classification of the orbit. Being more precise, blue colour corresponds to non-escaping regular orbits, magenta colour corresponds to trapped sticky orbits, yellow colour corresponds to trapped chaotic orbits, green colour corresponds to orbits escaping through $L_2$, while the initial conditions of orbits that escape through $L_3$ are marked with red colour. The outermost black solid line denotes the zero velocity curve, which is defined as
\begin{equation}
f_1(x,z) = \Phi_{\rm eff}(x, y = 0, z) = E.
\label{zvc}
\end{equation}

In panel (a) of Fig. \ref{xz} it is seen that for $E = -3240$, which is an energy level just above the energy of escape $E(L_2)$, about 90\% of the $(x,z)$ plane is covered by initial conditions of trapped chaotic orbits. The phenomenon of trapped chaos, for energy levels above yet very close to the energy of escape, has also been observed in other types of 3-dof systems, such as star clusters \citep[e.g.,][]{Z15,ZJ17} or planetary systems \citep[e.g.,][]{Z17}. Nevertheless, inside the unified trapped chaotic sea one can identify isolated initial conditions which correspond to escaping orbits. However, it is clear that for such low energy levels very close to the energy of escape a star is almost impossible to escape within a realistic time interval. As we proceed to higher values of the total orbital energy we observe that several basins of escape\footnote{A basin of escape is defined as a local set of initial conditions of orbits that escape through a certain escape channel.} start to emerge, mainly near the index-1 saddle points. At the same time, the portion of the trapped chaotic orbits is constantly been reduced, and when $E > -2890$ the corresponding initial conditions are mainly located in the vicinity of the boundaries of the stability islands. Furthermore, it is interesting to note that for $E > -3000$ the central areas of the $(x,z)$ plane around the main stability islands are covered by a highly fractal mixture of trapped chaotic and escaping orbits. At this point, it should be emphasized that when it is stated that an area is fractal we simply mean that it has a fractal-like geometry, without conducting any specific calculations regarding the fractal dimension as in \citet{AVS01}. In these fractal-like domains a high dependence of the final state on the particular initial conditions of the orbits is observed. This means that if we slightly perturb the $(x_0,z_0)$ initial conditions of an orbit the star will escape through the opposite escape channel. Of course this behavior is a classical indication of chaotic motion.

The following Fig. \ref{xzt}(a-f) illustrates the corresponding distribution of the escape time $t_{\rm esc}$ of the orbits presented in the CCDs of Fig. \ref{xz}(a-f). Again, we use a colour code in order to demonstrate the scale of the escape time. In particular, initial conditions of fast escaping orbits are indicated with light reddish colors, while initial conditions of orbits with high escape times are indicated by dark blue/purple colours. Moreover, white colour is used for non-escaping regular orbits and also for trapped sticky and chaotic orbits. It is interesting to note that orbits with initial conditions inside the basins of escape possess the shortest escape rates corresponding to no more than 500 time units. On the contrary, orbits with initial conditions inside the central fractal areas (around the boundaries of the stability islands) have escape periods which correspond to thousands of time units.

\begin{figure}
\begin{center}
\includegraphics[width=\hsize]{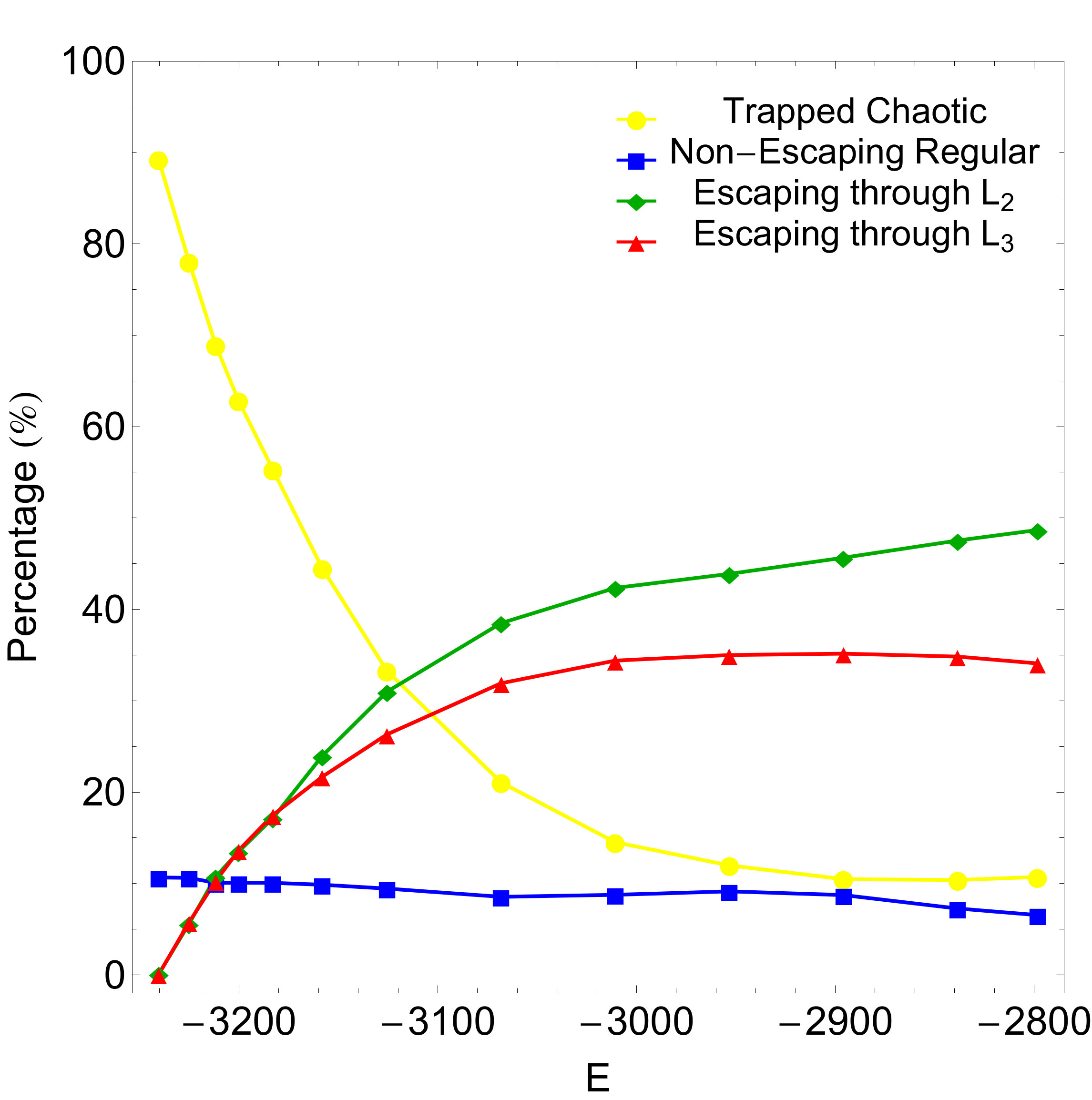}
\end{center}
\caption{Evolution of the percentages of all types of orbits on the $(x,z)$ plane, as a function of the total orbital energy $E$. The distribution shown applies to a particular plane of initial conditions. The asymmetry between the index-1 saddles $L_2$ and $L_3$ is caused by the particular choice of one branch of $p_{y0}$ (i.e. one orientation of the initial intersection of the plane $y = 0$). (For the interpretation of references to colour in this figure caption and the corresponding text, the reader is referred to the electronic version of the article.)}
\label{percs}
\end{figure}

\begin{figure}
\begin{center}
\includegraphics[width=\hsize]{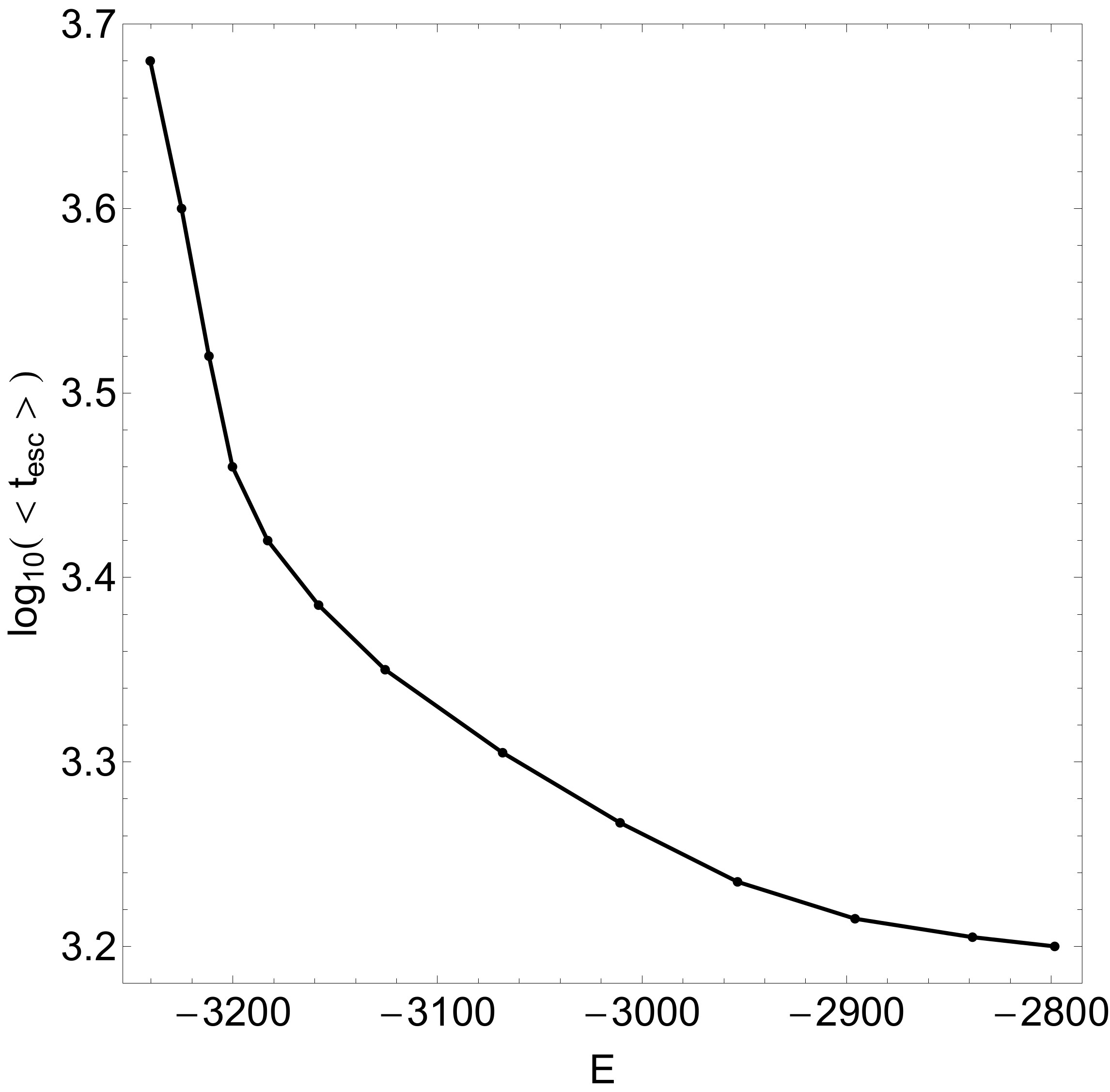}
\end{center}
\caption{Evolution of the logarithm of the average escape time of the orbits $(\log_{10}\left( < t_{\rm esc} > \right))$, as a function of the total orbital energy $E$.}
\label{tesc}
\end{figure}

\begin{figure*}
\centering
\resizebox{0.80\hsize}{!}{\includegraphics{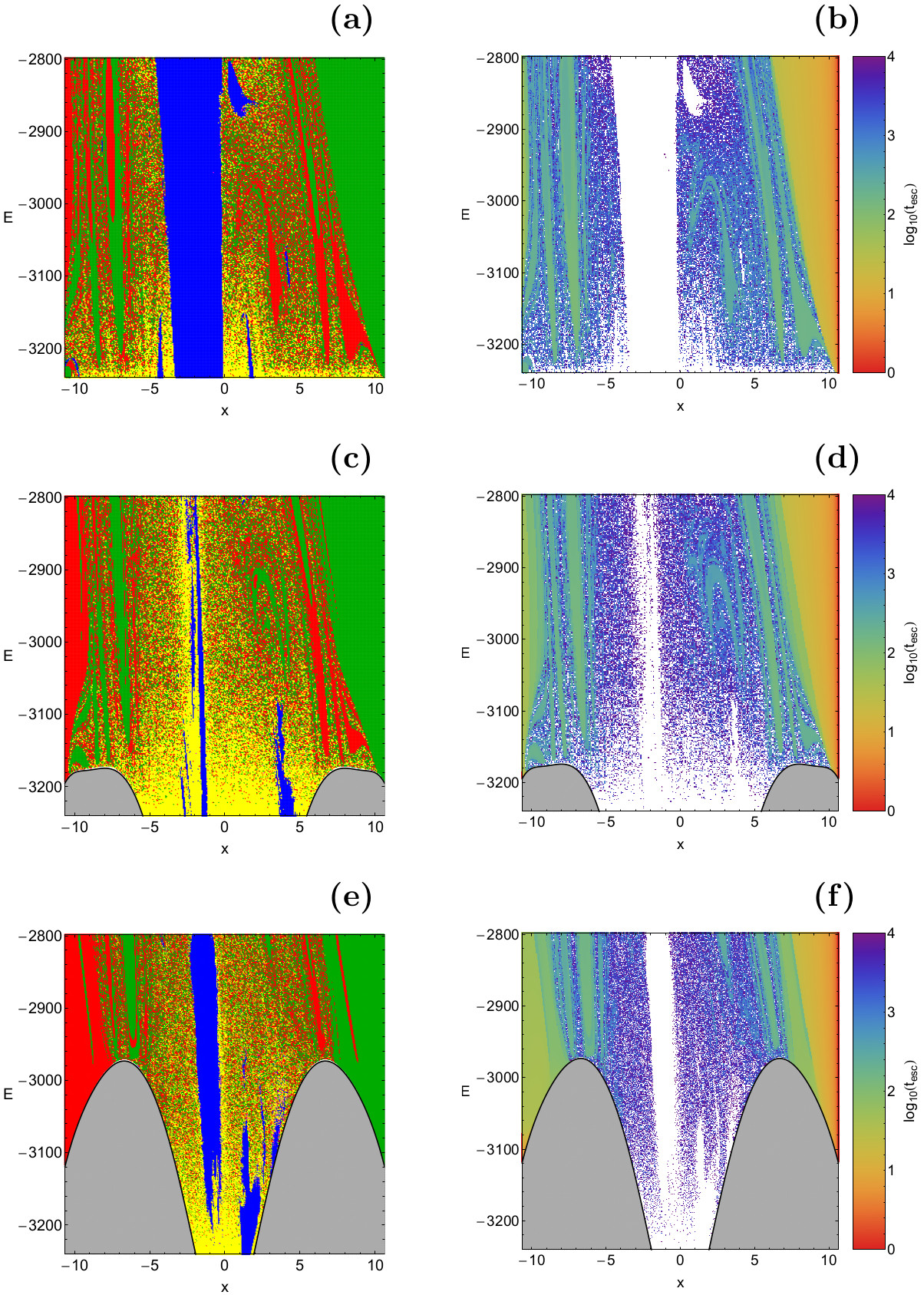}}
\caption{(left column): Orbital structure of the $(x,E)$ plane when (a): $z_0 = 0.01$; (b): $z_0 = 1$; (c): $z_0 = 2$. The colour code is as follows: non-escaping regular orbits (blue), trapped sticky orbits (magenta), trapped chaotic orbits (yellow), escaping orbits through $L_2$ (green), escaping orbits through $L_3$ (red). The energetically forbidden areas are marked with grey. (right column): Distribution of the corresponding escape time $t_{\rm esc}$ of the orbits on the $(x,E)$ plane for the corresponding values of $z_0$. (For the interpretation of references to colour in this figure caption and the corresponding text, the reader is referred to the electronic version of the article.)}
\label{xEt}
\end{figure*}

\begin{figure*}
\centering
\resizebox{0.80\hsize}{!}{\includegraphics{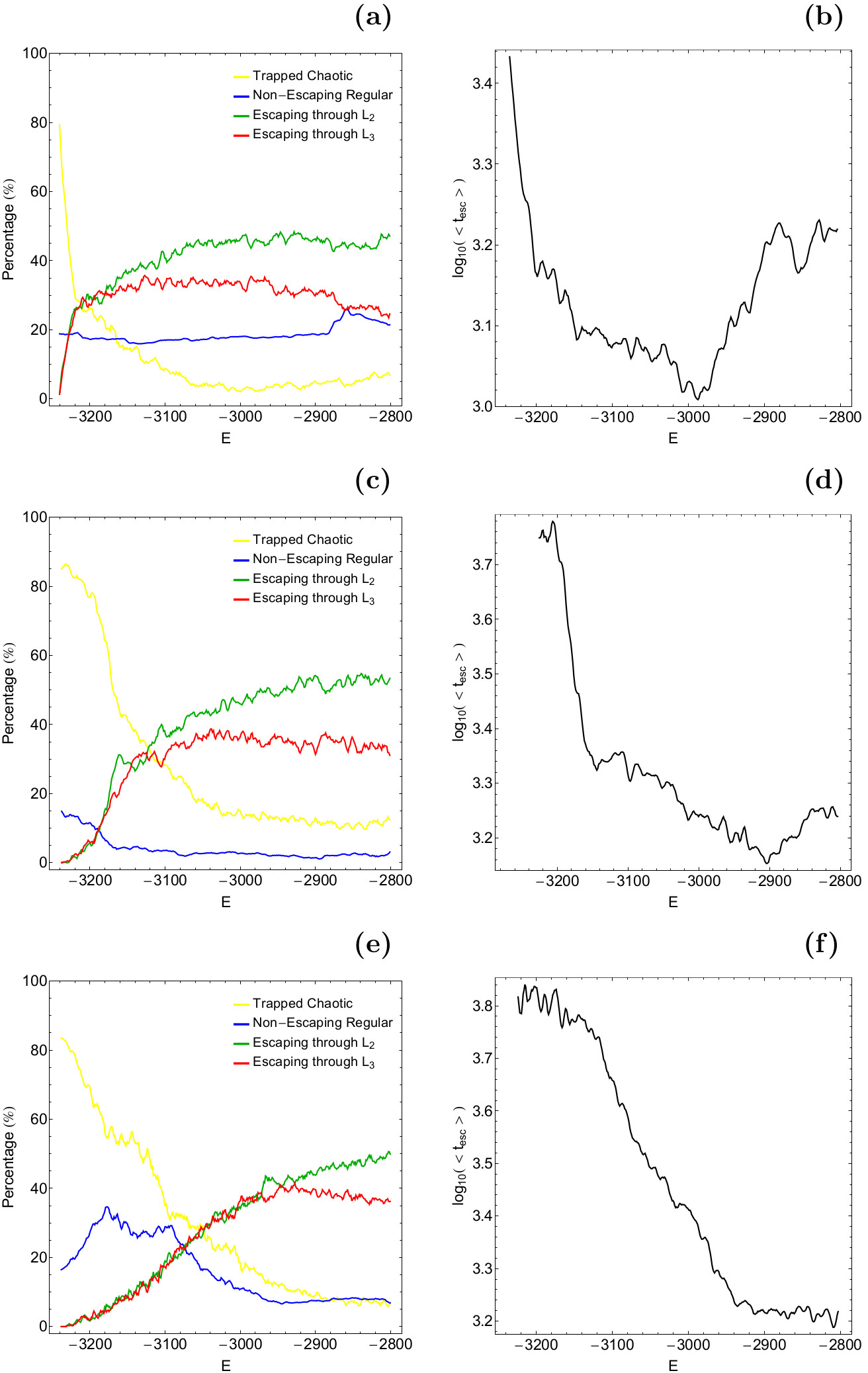}}
\caption{Evolution of the (left column): percentages of all types of orbits and (right column): logarithm of the average escape time of the orbits $(\log_{10}\left( < t_{\rm esc} > \right))$ on the $(x,E)$ plane, as a function of the total orbital energy $E$. (first row): $z_0 = 0.01$; (second row): $z_0 = 1$; (third row): $z_0 = 2$. (For the interpretation of references to colour in this figure caption and the corresponding text, the reader is referred to the electronic version of the article.)}
\label{stats}
\end{figure*}

Important information, regarding the orbital structure of the system, can be extracted if we monitor the evolution of the percentages of all types of orbits, as a function of the total energy $E$. Such a diagram is given in Fig. \ref{percs}. Here we would like to clarify that for constructing this diagram we exploited data from additional CCDs, apart from those presented earlier in Fig. \ref{xz}. We see that just above the energy of escape trapped chaotic orbits dominate by covering more than 90\% of the $(x,z)$ plane. However as the value of the energy increases the rate of trapped chaotic orbits smoothly decreases and for $E > -2900$ it saturates at about 10\%. The percentage of non-escaping regular orbits seems to be almost unperturbed by the change of the value of the total orbital energy. Indeed our calculations indicate that when $E \in (E(L_2), E(L_4)]$ non-escaping regular orbits have a mean rate of 12\%. Looking at Fig. \ref{percs} one can deduce that for values of energy very close to the escape energy the percentages of both escape channels are equal which of course means that both exits are equiprobable. On the other hand, for $E > -3180$ the rates of escaping orbits start to diverge. Specifically, the amount of escaping orbits through $L_2$ increases more rapidly with respect to the portion of escaping orbit through $L_3$. At the highest energy level studied, that is $E = E(L_4)$, escaping through $L_2$ occupies about half of the $(x,z)$ plane, while initial conditions of orbits that escape over $L_3$ is about 35\%. Thus, taking into account all the above-mentioned analysis one may conclude that at relatively low energy levels, where the fractality of the $(x,z)$  plane is maximum, stars do not show any particular preference regarding the escape channels. On the contrary, at high enough energy levels, where basins of escape dominate, it seems that escape over $L_2$ is more preferable with respect to escape over $L_3$.

At this point we would like to point out that our numerical computations suggest that sticky orbits always possess an extremely low percentage (less than 0.001\%), which strongly implies that the chosen total time of the numerical integration $(t_{\rm max} = 10^4)$ is indeed sufficient, so as almost all sticky orbits to be correctly classified as chaotic ones.

Of course, the relative fraction of orbits escaping through index-1 saddles $L_2$ and $L_3$ strongly depends on our particular choice of the collection of initial conditions. The inversion symmetry of the whole system implies that to each orbit escaping through saddle $L_2$ there is another symmetric orbit, with inverted initial conditions, which escapes through the saddle $L_3$. Therefore an integration over the whole 5-dimensional energy shell of possible initial conditions would give equal escape rates through both saddles. In addition, a random collection of initial conditions from the interior part of the 5-dimensional energy shell would lead to equal escape percentages. However, for sets of initial conditions on any lower dimensional surface these escape rates depend on how this particular surface intersects the basins of escape belonging to the two index-1 saddle points.

Earlier in Fig. \ref{xzt} we have seen that with increasing value of the total orbital energy $E$ the escape period of the orbits is reduced. In Fig. \ref{tesc} we present the evolution of the average value of the escape time $< t_{\rm esc} >$ of the orbits as a function of $E$. It is seen that just above the energy of escape the average escape time of the stars correspond to about 4800 time units. However, as the value of the energy grows the required time for escape smoothly reduced until $E = E(L_4)$, where $< t_{\rm esc} > \simeq 1500$ time units. The observed behavior can be explained as follows: as we proceed to higher energy levels the two escape channels, around the index-1 saddle points, become more and more wide. Therefore, the stars need less and less time in order to find one of the two symmetrical exits and eventually escape from the Lagrange radius. This simple geometrical feature explains, in a very satisfactory way, the parametric evolution of the escape time, shown in Fig. \ref{tesc}.

The CCDs presented in Fig. \ref{xz} can provide useful information, regarding the orbital structure of the $(x,z)$ plane, however only for some specific values of the total orbital energy $E$. In order to overcome this limitation and therefore obtain a more complete view of the orbital structure of the dynamical system, we shall adopt the H\'{e}non's method \citep{H69}, thus trying to gather information of a continuous spectrum of energy values. In particular, for specific values of $z_0$ we define dense uniform grids of initial conditions on the $(x,E)$ plane with $y_0 = p_{x_0} = p_{z_0} = 0$, while in all cases the initial value of $p_y$ is obtained from the Jacobi integral (\ref{ham}). Using this method we are able to monitor the evolution of the orbital structure of the system, using a continuous spectrum of energy values, rather than a few discrete ones. The orbital structure of the $(x,E)$ plane, when $E \in (E(L_2),E(L_4)]$, along with the distribution of the corresponding escape time of the orbits, is presented in Fig. \ref{xEt}(a-f). The black solid line is the limiting curve which in this case is defined as
\begin{equation}
f_2(x,E;z_0) = \Phi_{\rm eff}(x,y = 0,z = z_0) = E.
\label{zvc2}
\end{equation}

Panels (a) and (b) of Fig. \ref{xEt} correspond to $z_0 = 0.01$, that is a low value of initial coordinate $z$. In other words this means that the three dimensional orbits are started very close to the galactic $(x,y)$ plane. We observe that throughout the energy range the stability islands of simple loop orbits, that are parallel to the galactic plane, are present near the central region, while the rest of the $(x,E)$ plane is covered either by well-formed basins of escape or by highly fractal domains. When $z_0 = 1$ we see in panels (c) and (d) of Fig. \ref{xEt} that bounded basins of non-escaping regular motion are very limited. Moreover, the majority of the $(x,E)$ plane displays a very complicated structure with multiple basins of escape, surrounded by highly fractal regions. Finally, for $z_0 = 2$ (see panels (e) and (f) of Fig. \ref{xEt}) the presence of non-escaping regular motion is strong again. However this time the elongated stability island for $x < 0$ is composed of initial conditions that correspond to titled loop orbits. It is interesting to note that as the initial value of the $z$ coordinate increases the energetically forbidden regions grow rapidly, at both sides of the $(x,E)$ plane.

The evolution of the percentages of all types of orbits with initial conditions on the $(x,E)$ planes of Fig. \ref{xEt} are given in the left column of Fig. \ref{stats}. There we observe a very interesting phenomenon. In Fig. \ref{percs}, which corresponds to data from the $(x,z)$ planes, we seen that the percentages of both escape channels are almost identical when $E(L_2) < E < -3180$, while for higher values of the energy the two rates start to diverge. For the data of the $(x,E)$ planes the energy point where the diversion of the percentages start depend on the value of $z_0$. More specifically, the energy diversion points are -3220, -3180, and -2970 when $z_0 =$ 0.01, 1, and 2, respectively. However in all studied cases for relatively high values of the energy the amount of escaping orbits through $L_2$ are always higher than that of escaping orbits through $L_3$.

In the right column of Fig. \ref{stats} we present the corresponding evolution of the average value of the escape time of the orbits. It is seen that in all cases the average escape period starts at very high values, while its value is reduced, with increasing value of the total orbital energy. Nevertheless, when $z_0$ is equal to 0.01 and 1 one can see that after a certain value of the energy the average escape period starts to increase. A possible explanation for this phenomenon is the following: in this cases some secondary basins of escape disappear at certain values of $E$ thus giving space to fractal regions where the escape time of the orbits is higher.

Before closing this section we would like to mention that for the numerical integration of each set of the initial conditions of the orbits, in all types of colour-coded grid presented here, we needed roughly about 2.8 days of CPU time on an Intel$^{\circledR}$ Quad-Core i7 2.4 GHz PC.

\section{Orbital dynamics}
\label{orbdyn}

\begin{figure*}
\centering
\resizebox{\hsize}{!}{\includegraphics{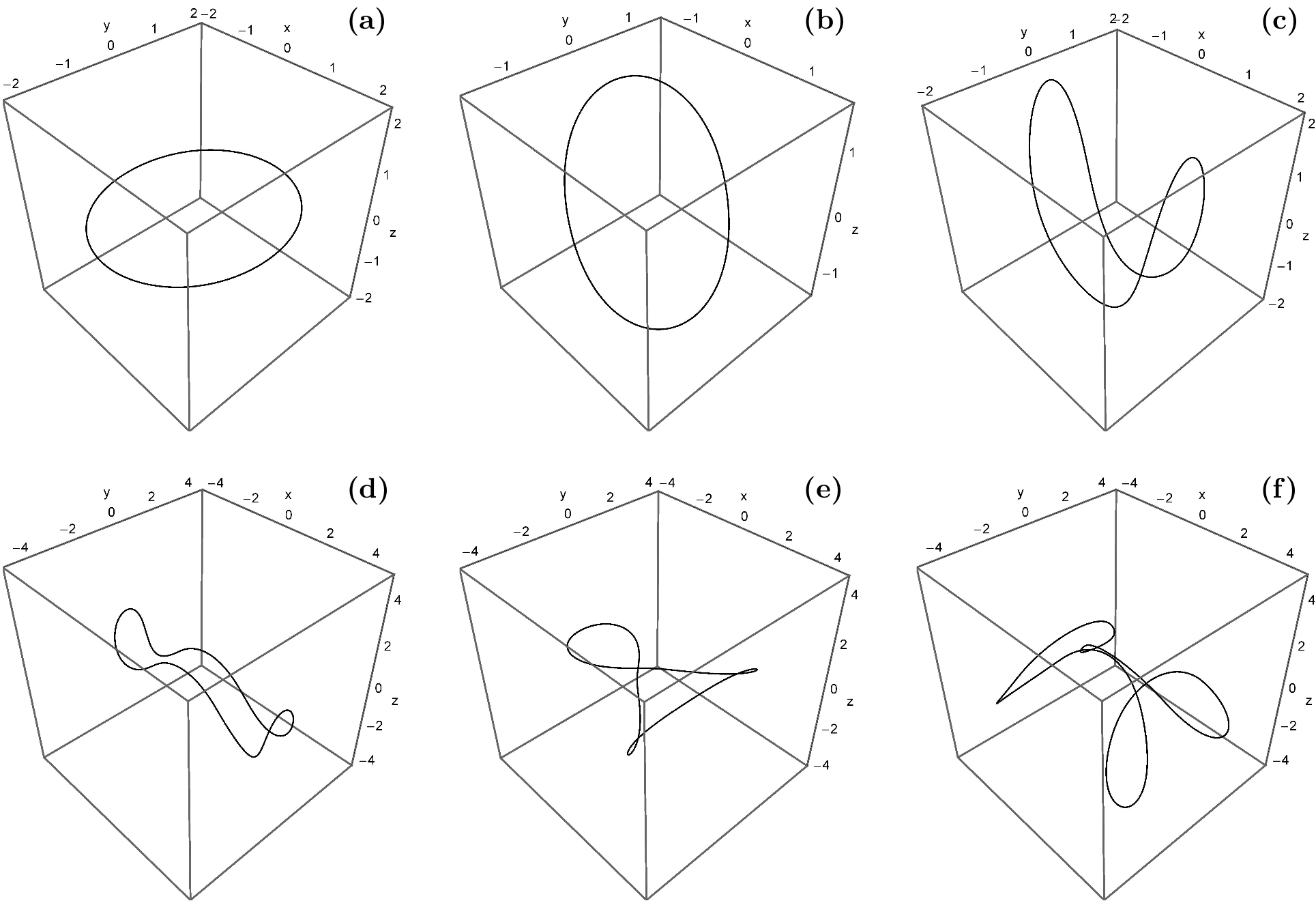}}
\caption{A collection of the main types of orbital families observed in our barred galaxy model. The corresponding initial conditions of the parent periodic orbits, as well as the respective values of the energy, are given in Table \ref{tab1}.}
\label{orbs}
\end{figure*}

\begin{figure}
\begin{center}
\includegraphics[width=\hsize]{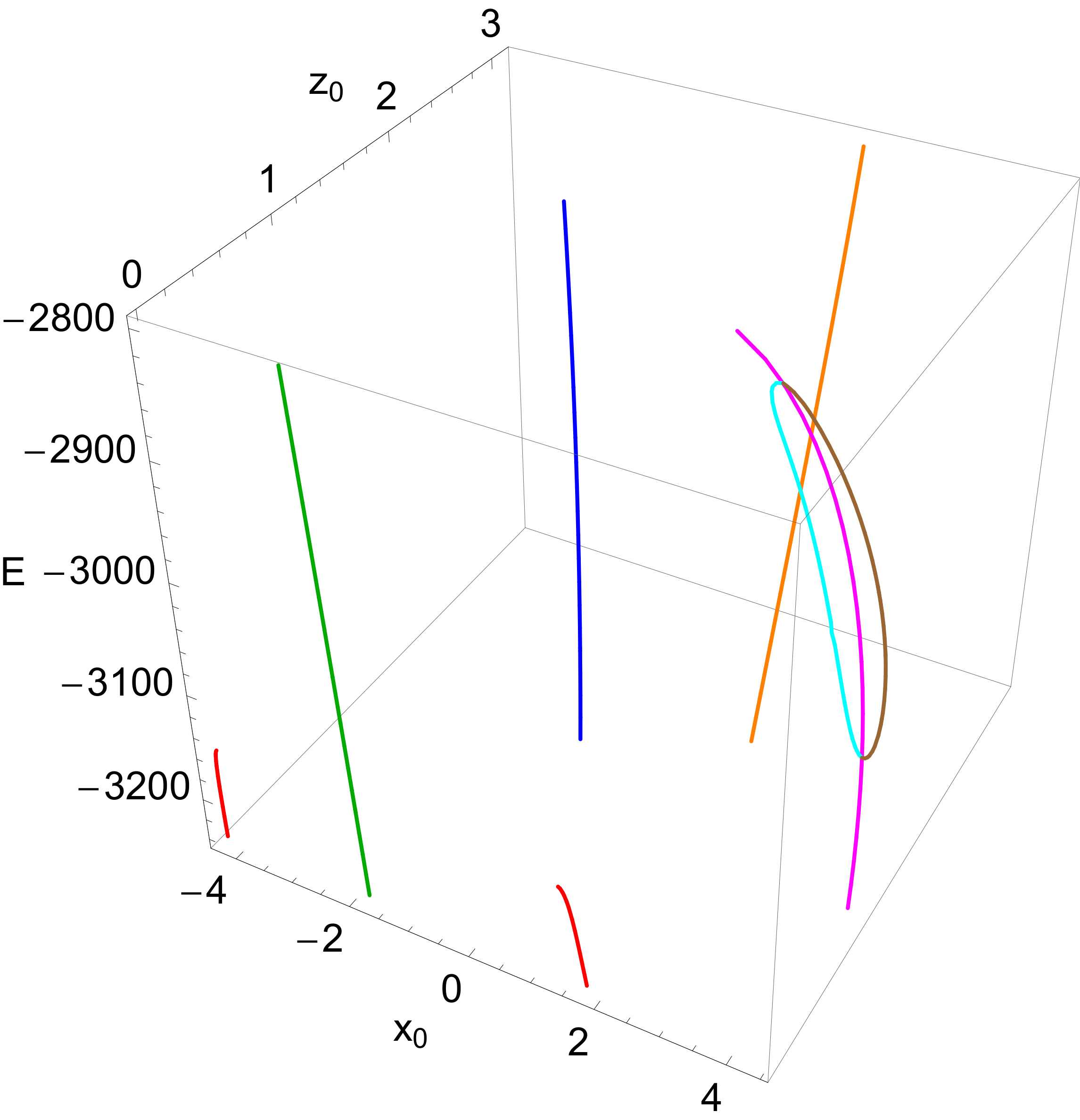}
\end{center}
\caption{Evolution of the $(x_0,z_0)$ initial conditions of the parent periodic orbits, as a function of the total orbital energy $E$. The colour code is as follows: $f_1$ family (green); $f_2$ family (blue); $f_3$ family (orange); $f_4$ family (magenta); $f_5$ family (red); first bifurcated family related to $f_4$ (cyan); second bifurcated family related to $f_4$ (brown). (For the interpretation of references to colour in this figure caption and the corresponding text, the reader is referred to the electronic version of the article.)}
\label{fpo}
\end{figure}

In CCDs of Fig. \ref{xz} we have seen that a considerable amount of the $(x,z)$ plane is covered by initial conditions corresponding to non-escaping regular orbits. We observe that throughout the energy interval $(E(L_2), E(L_4)]$ regular motion is always possible, while the main structure of the stability islands remain almost unperturbed by the shift of the total orbital energy $E$. Additional numerical calculations indicate that the observed stability islands correspond to a plethora of regular orbits. In Fig. \ref{orbs} we present a collection of the parent periodic orbits of the most important orbital families that exist in our barred galaxy model, while in Table \ref{tab1} we provide the exact initial conditions and the classification of the orbits. Our analysis suggests that the vast majority of the non-escaping initial conditions correspond to simple loop orbits which are either parallel to the galactic $(x,y)$ plane (see e.g., panel (a) of Fig. \ref{orbs}) or tilted with respect to it (see e.g., panel (b) of Fig. \ref{orbs}). Nevertheless, orbits of higher resonances are also present (see e.g., panel (f) of Fig. \ref{orbs}), although their rate is very small, compared to that of the simple loop orbits. Moreover, regular orbits of higher resonances are mainly observed for values of energy above yet very close to the energy of escape.

\begin{table}
 \center
 \caption{The $(x_0,z_0)$ initial conditions and the value of the energy $E$ of the periodic orbits presented in the six panels of Fig. \ref{orbs}. All orbits have $y_0 = p_{x0} = p_{z0} = 0$, while the value of $p_{y0}$ is obtained from the Jacobi integral of motion (\ref{ham}).}
 \label{tab1}
 \begin{tabular}{lccccc}
  \hline
  Panel & $E_0$ & $x_0$ & $z_0$ & Classification & Family\\
  \hline\hline
  (a) & -3240    & -1.7627 & 0.0000 & 1:1:0 & $f_1$ \\
  (b) & -3240    & -0.7524 & 1.5425 & 1:1:1 & $f_2$ \\
  (c) & -3240    &  1.5958 & 1.9770 & 1:1:2 & $f_3$ \\
  (d) & -3240    &  4.3316 & 1.0773 & 1:1:4 & $f_4$ \\
  (e) & -3240    &  1.7894 & 0.0000 & 1:2:0 & $f_5$ \\
  (f) & $E(L_4)$ &  2.1490 & 1.1894 & 1:1:4 & $f_4$ \\
  \hline
 \end{tabular}
\end{table}

The integers $n$, $m$, $l$, explaining the classification $n:m:l$ (see Table \ref{tab1}) are the numbers of switches of the sign of the coordinates from negative to positive. In particular, we count full oscillations, therefore a full oscillation means 2 changes of the sign. This method works well only for orbits that are symmetric in the position space.

So far we seen that there are five main orbital families, $f_i$, $i=1,..., 5$ in our barred galaxy model. Therefore it would be very informative to reveal how the corresponding parent periodic orbits evolve as a function of the total orbital energy. Our results are presented in Fig. \ref{fpo}, where we provide the parametric evolution of the $(x_0, z_0)$ initial conditions of the periodic orbits in the energy interval $E \in (E(L_2), E(L_4)]$. For all orbits $y_0 = p_{x0} = p_{z0} = 0$, while the initial value of the momenta $p_y$ is derived from the energy integral (\ref{ham}). It is interesting to note that additional families of the periodic orbits bifurcate from the orbital family $f_4$.

\begin{figure}
\begin{center}
\includegraphics[width=\hsize]{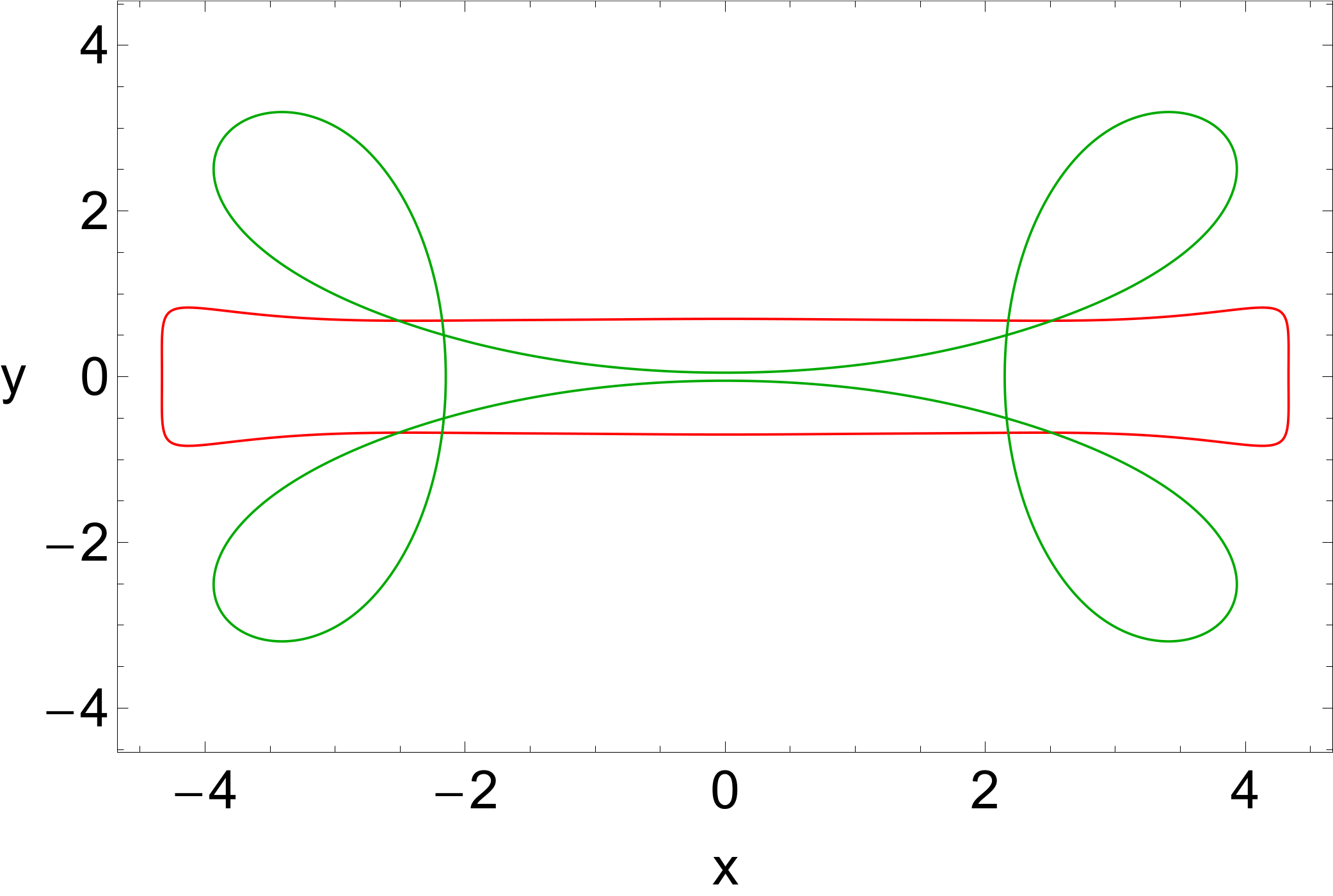}
\end{center}
\caption{Projection on the galactic $(x,y)$ plane of the x1 orbits, when $E = -3240$ (red) and $E = E(L_4)$ (green). (For the interpretation of references to colour in this figure caption and the corresponding text, the reader is referred to the electronic version of the article.)}
\label{x10}
\end{figure}

\begin{figure}
\begin{center}
\includegraphics[width=\hsize]{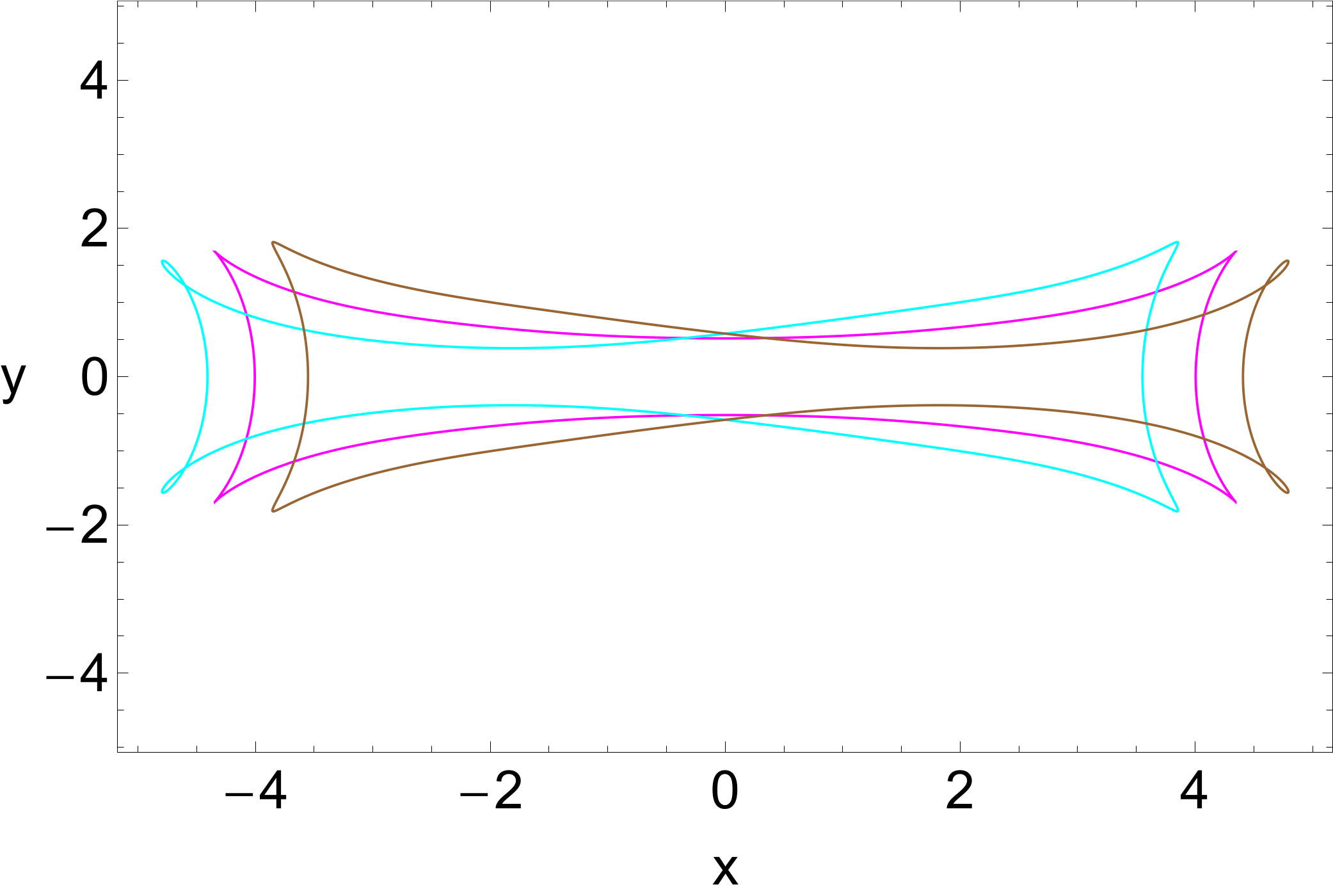}
\end{center}
\caption{The unstable central periodic orbit (magenta), along with the two split off periodic orbits (cyan and brown), when $E = -3040$. (For the interpretation of references to colour in this figure caption and the corresponding text, the reader is referred to the electronic version of the article.)}
\label{x1}
\end{figure}

\begin{figure}
\begin{center}
\includegraphics[width=\hsize]{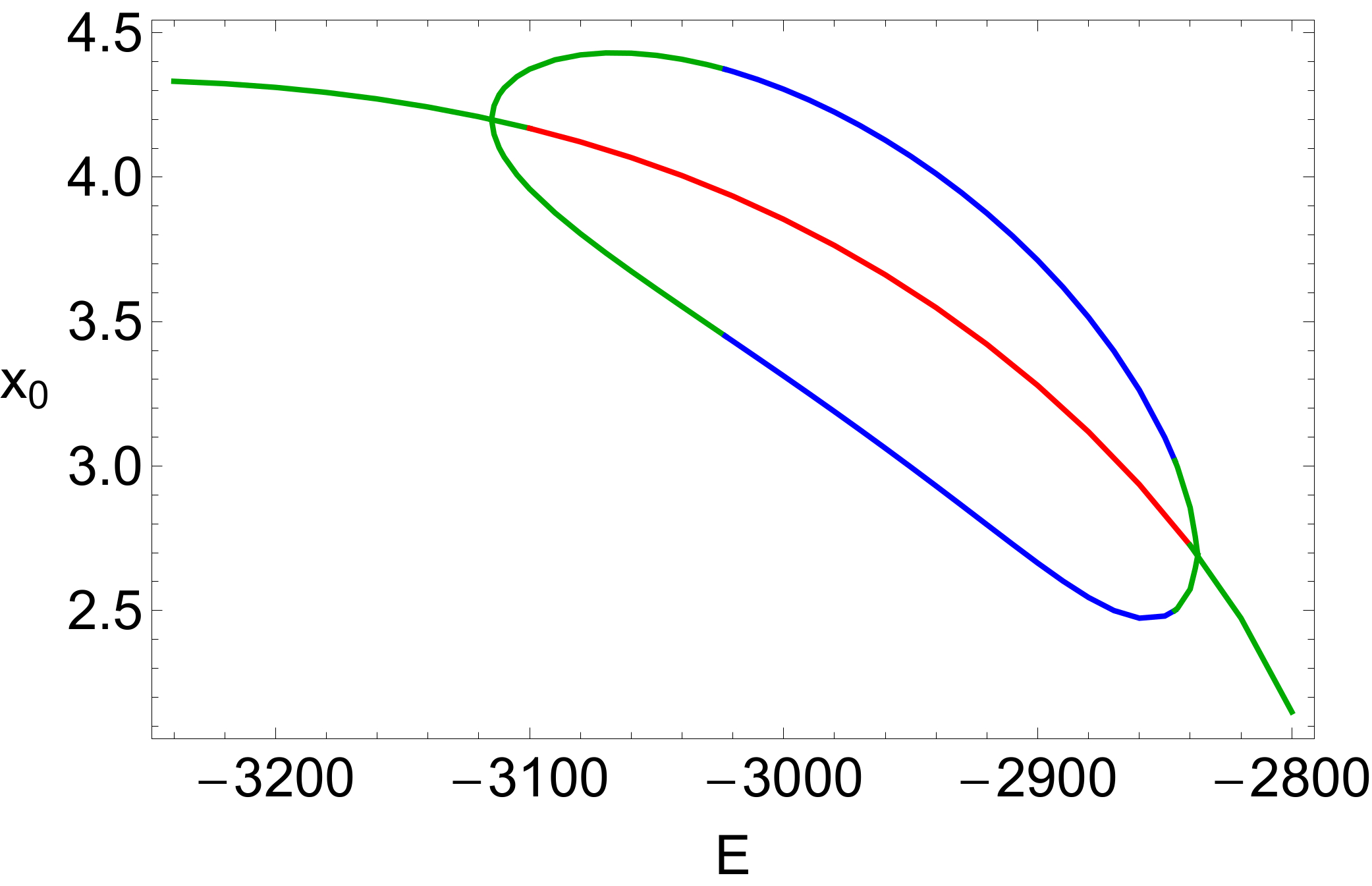}
\end{center}
\caption{Evolution of the stability of the x1 periodic orbits, as a function of the total orbital energy $E$. Green colour indicates stable motion, red colour indicates unstable motion, while blue colour corresponds to the complex spiral case, where the monodromy matrix cannot be decomposed into two real $ 2 \times 2$ blocks. (For the interpretation of references to colour in this figure caption and the corresponding text, the reader is referred to the electronic version of the article.)}
\label{stab}
\end{figure}

\begin{figure*}
\centering
\resizebox{0.75\hsize}{!}{\includegraphics{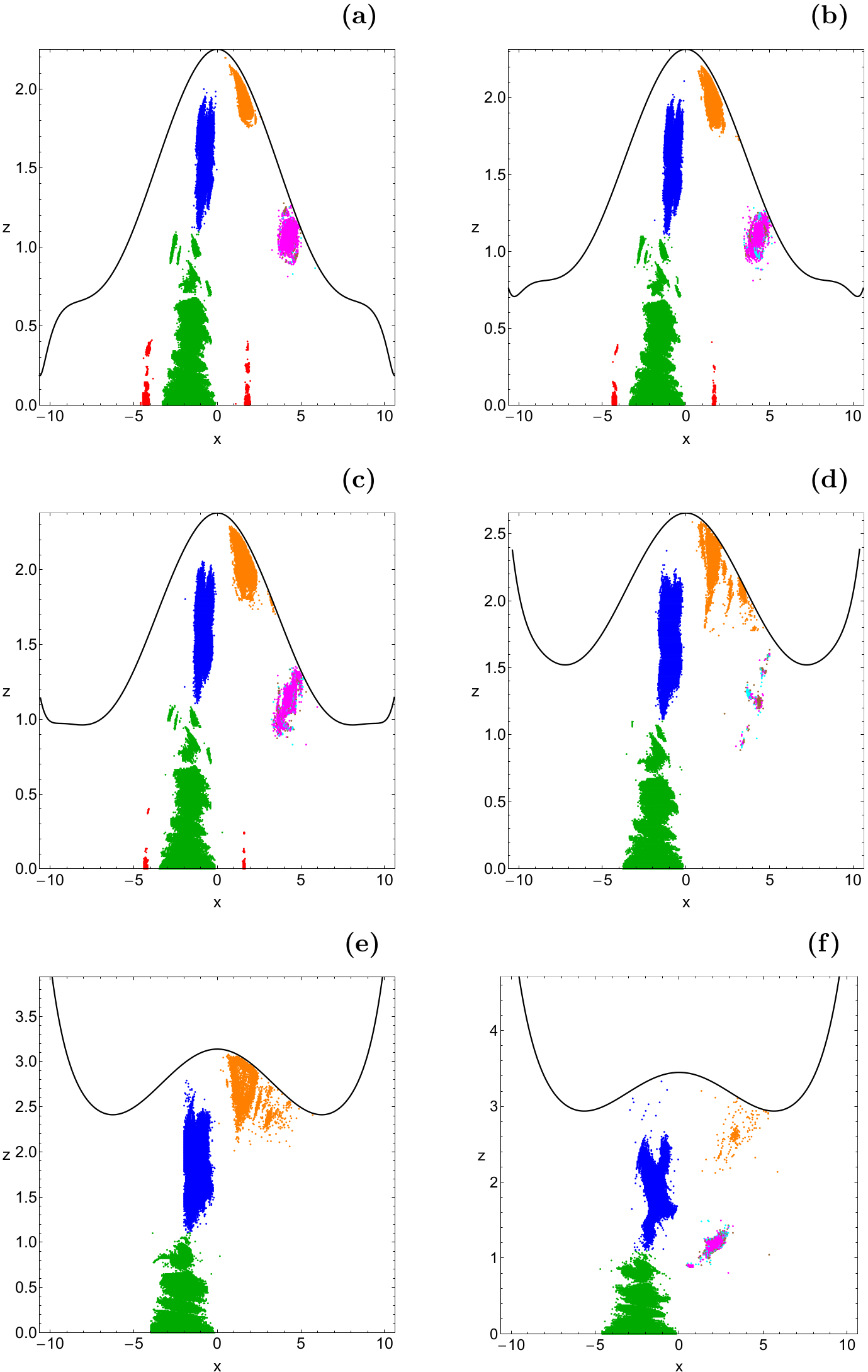}}
\caption{Classification of the regular orbits on the $(x,z)$ plane into the main orbital families. The energy levels in each panel are the same with the corresponding ones of Fig. \ref{xz}(a-f). The colour code is as follows: $f_1$ family (green); $f_2$ family (blue); $f_3$ family (orange); $f_4$ family (magenta); $f_5$ family (red); first bifurcated family related to $f_4$ (cyan); second bifurcated family related to $f_4$ (brown). (For the interpretation of references to colour in this figure caption and the corresponding text, the reader is referred to the electronic version of the article.)}
\label{reg}
\end{figure*}

The shapes of bars are dynamically stabilized by bundles of orbits which, in the rotating frame, oscillate essentially along the bar, while making smaller wiggles in the transverse directions. As has been mentioned in subsection 3.1 of \citep{ARGBM09} the projections of the bars into the horizontal $(x,y)$ plane many times have approximate rectangular shapes. And this shape could be explained by stable x1 orbits having this rectangular shape and forming strong bundles of similar orbits around them. As claimed in \citet{ARGBM09} such periodic orbits are not generally known. The reason for this might be that people usually search for in plane orbits using a 2-dof dynamics of this in plane motion only. In contrast we use the full 3-dof dynamics and interestingly we find periodic orbits of this rectangular shape in their projection into the horizontal galactic plane. The most simple and central one is the orbital family $f_4$, already presented in panel (d) of Fig. \ref{orbs}. In Fig. \ref{x10} we plot its projection into the horizontal plane together with the analogous projection of the orbit from panel (f) of Fig. \ref{orbs}, which is the same orbit at the higher energy $E(L_4)$. We see that this projection of the orbit at energy $E = -3240$ is almost a perfect rectangle with the long direction oriented along the bar direction. Because of symmetry reasons there is also the corresponding $z$-reflected orbit which of course has the same projection into the galactic plane.

The development scenario of the x1 orbital family is roughly the following: At energies way below the saddle energy $E(L_2)$ it encircles the minimum of the potential in the origin and when its energy approaches the saddle energy from below then its form approaches the rectangle seen in the plot of Fig. \ref{x10}. At such energies it is dynamically stable. And this orbit is always left right symmetric along the $x$-direction. At the higher energy $E = -3115$ it becomes unstable in $x$-direction and in a pitchfork bifurcation it splits off a pair of similarly looking orbits which individually break the left right symmetry in the $x$-direction. However, the global $x$ symmetry of the system is conserved in the sense that one of these split off orbits is the $x$ reflected image of the other one. In the pitchfork bifurcation the original x1 orbit becomes unstable but the two split off orbits are dynamically stable, up to the energy -3024, and take over the role of the guiding centers of the whole bundle of similar orbits. The orbits in this bundle are partly stable and partly weakly unstable. The now unstable original orbit and the two split off orbits at the energy $E = -3040$ are plotted in figure \ref{x1}.

In the next plot of Fig. \ref{stab} we provide the evolution of the stability properties of all the x1 periodic orbits (the main $f_4$ family plus the two bifurcated families), as a function of the energy $E$. First we calculated the 4-dimensional monodromy matrix of these orbits, then whenever it is possible we decompose them into two real 2-dimensional blocks. Our analysis suggests that the orbital families $f_1$ and $f_2$ are always stable, while $f_3$ is stable only when $E(L_2) < E \leq -2940$ and $f_5$ is stable only when $E(L_2) < E \leq -3156$.

At the still higher energy $E = -2837$ the symmetric central x1 orbit becomes dynamically stable again. However at such high energies close to $E(L_4)$ the motion along the $y$ and the $z$ directions is already so wide that the orbit no longer qualifies as a valid x1 orbit (see e.g., panel (f) of Fig. \ref{orbs}). Therefore we may argue that x1 orbits are being deformed at relatively high energy levels. However, such high energy levels are no longer relevant neither for the saddle escape nor for the stable motion inside of the bar.

In Fig. \ref{reg}(a-f) we illustrate how the different orbital families are distributed on the $(x,z)$ plane, for the same energy levels of Fig. \ref{xz}(a-f). For producing this figure we worked as follows: first we extracted from the data corresponding to the $(x,z)$ planes of Fig. \ref{xz} all the initial conditions of the non-escaping regular orbits. Then using the numerical algorithm Taxon, introduced in \citet{CA98} and upgraded in \citet{ZC13}, we reclassified all the regular orbits into the main orbital families. According to panel (e) of Fig. \ref{reg} when $E = -2895$ there is no numerical indication of x1 orbits, while when they reappear for $E = E(L_4)$ (see panel (f)) they have already lost the x1 property. Panel (d) of Fig. \ref{reg} shows the regions of stable motion for the energy level $E = -3068$, where the original x1 orbits is already unstable, but the split off orbits are still stable. Accordingly, the point belonging to the original x1 orbit lies in a small white gap between the most important cyan and brown regions, whose centres are the two split off orbits, respectively. This is the form how the CCDs show graphically how the split off orbits take over the stability and become the new organization centres of the bundle of stable motion, belonging to the $f_4$ family of orbits.

In \citetalias{JZ16a} we have seen that in the 2-dof system several types of x1 orbits exist (see Fig. 5 in \citetalias{JZ16a}), however all these planar orbits are highly unstable. The fact that we find stable x1 orbits only in the full 3-dof system might have the following explanation. For a 2-dof calculation all energy has to be put into the $x$ and $y$ degrees of freedom and then there is too much energy in the dynamics to allow the formation of stable in plane periodic orbits of the desired rectangular shape. However, when we treat the full 3-dof dynamics then a part of the total energy goes into the $z$ degree of freedom and only a correspondingly smaller amount of the energy is left for the in plane motion. And at this smaller amount of in plane energy it seems to be possible to obtain the periodic orbits of approximate rectangular shape. With this idea in mind, note that already for the small energy $E = -3240$ the z oscillations are strong which means that a considerable part of the total energy has gone into the $z$ motion.

Finally we would like to report that for extremely high energy levels, way above the saddle energy $E(L_4)$, only the most basic and simple types of orbits survive, that is the loop orbits which are parallel or tilted with respect to the galactic $(x,y)$ plane.

\section{Connection between basin boundaries and the NHIMs}
\label{bbs}

In Section \ref{escdyn} we have seen how the initial conditions chosen in any general plane in the phase space are sorted into various groups. These groups are: (i) escape through $L_2$, (ii) escape through $L_3$, (iii) regular bounded motion with no escape at all and finally (iv) escape only after an extremely long time of transient irregular bounded motion. On the other hand it is our claim that all qualitative features of the whole dynamics are essentially directed by the NHIMs, because they are the most important elements of the skeleton of the whole dynamics. To corroborate this claim we have to demonstrate in which sense the NHIMs are responsible for the division of the plane of initial conditions into the subregions belonging to the various types of behaviour.

\begin{figure*}
\centering
\resizebox{\hsize}{!}{\includegraphics{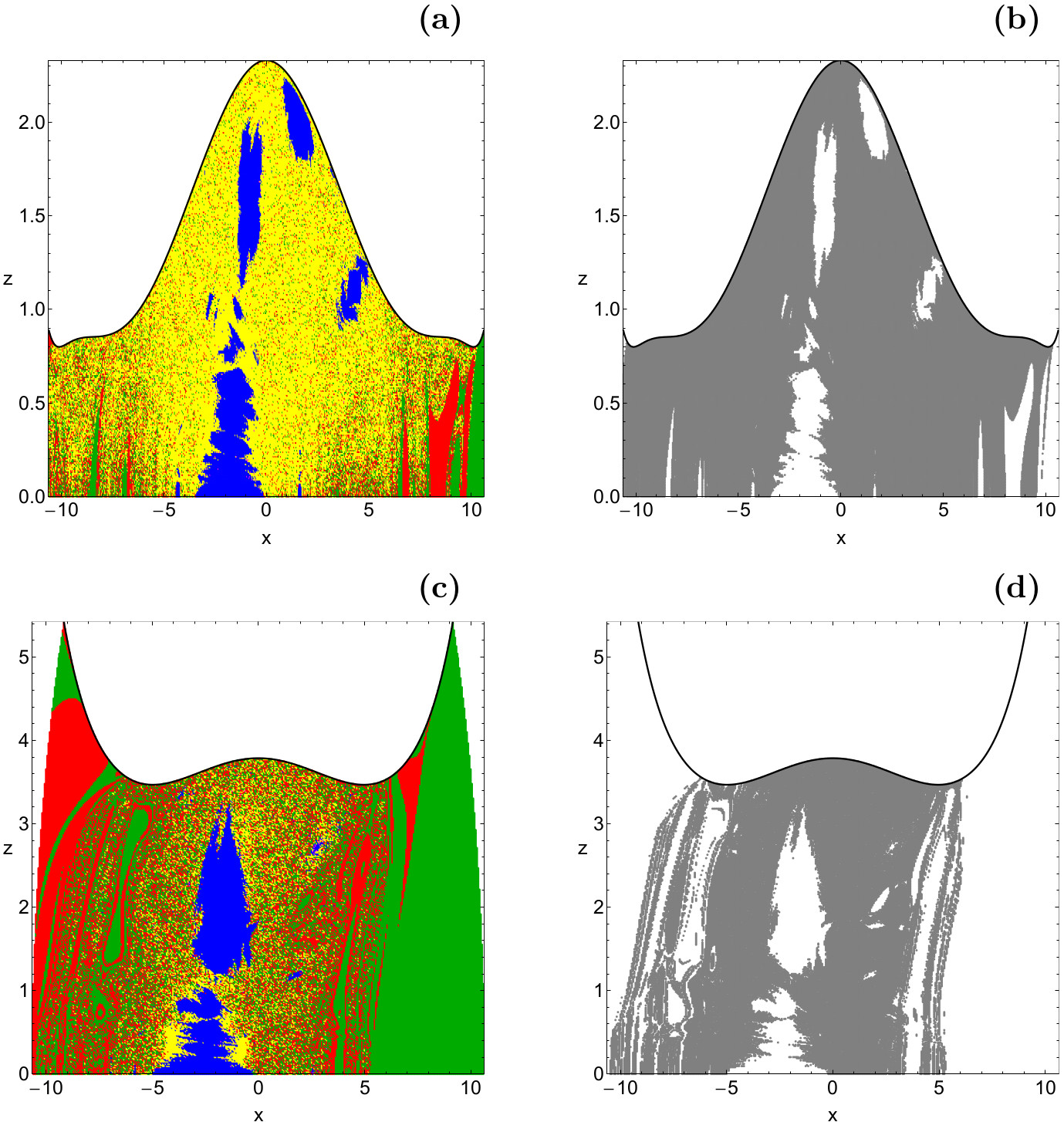}}
\caption{Comparison of the CCDs with the intersections of stable manifolds of saddle NHIMs with the domain of the CCDs. Panels (a) and (c) are the CCDs, explained in Section \ref{escdyn}, for the energy values -3200 and -2700, respectively. The panels (b) and (d) mark in the same domain the points close to intersections with the stable manifolds of the saddle NHIM, again for the energies -3200 and -2700, respectively. (For the interpretation of references to colour in this figure caption and the corresponding text, the reader is referred to the electronic version of the article.)}
\label{inter}
\end{figure*}

The essential idea is provided by a consideration of trajectories which start very close to the stable manifold of a NHIM. These trajectories approach the NHIM and thereby the corresponding potential saddle and then the further fate of the trajectory depends on which side of the local segment of the stable manifold it runs. Trajectories coming in on one side will pass the saddle and escape. Trajectories coming in on the other side will return to the interior region. Of course, trajectories which return to the inner potential well can later approach again one of the saddles and escape after several previous failed attempts. The important property is that the local segment of the stable manifold of the NHIM separates the phase space region of immediate escape from the phase space region of return. Thereby it is clear in which sense the stable manifold is a boundary between qualitatively different behaviour. It should be emphasised that we speak here about NHIMs over index-1 saddles, therefore the NHIMs are surfaces of codimension 2 in the phase space and also in the domain of the Poincar\'{e} map and accordingly their stable manifolds are of codimension 1 and really separate distinct sides, at least locally. Stable manifolds of NHIMs of higher codimension would themselves have higher codimension and would not be divisions in the full dimensional phase space or full dimensional domain
of the Poincar\'{e} map.

A general 2-dimensional surface in phase space is intersected transversally by the stable manifold in a collection of 1-dimensional lines on this surface. Globally the stable manifold is folded in a complicated, in general fractal form, and therefore also the set of intersection lines with the surface is a fractal collection of lines. So the basin boundaries on the surface are a complicated fractal set. This is exactly what has been seen in Fig. \ref{xz}. It should be emphasised that strictly speaking the yellow and the magenta points both belong to the types of trajectories which can escape after very long bound transients and accordingly all yellow and all magenta points lie close to some intersection with the stable manifold of some NHIM from one of the index-1 saddles.

To demonstrate this mechanism graphically we present Fig. \ref{inter}. Here panel (a) shows the CCD for $E = -3200$, while panel (c) shows the CCD for $E = -2700$. The first energy is one rather close to the saddle energy and the second one is close to the highest energy for which the NHIM over the potential saddle is still complete. For energies a little higher than -2700 the decay of the NHIM starts. To construct panels (b) and (d) of the figure, again for energies -3200 and -2700 respectively, the following procedure has been applied: $10^9$ points very close to the NHIM and well distributed over the NHIM have been chosen as initial points and the corresponding trajectories in phase space have been constructed and followed into the past. These past trajectories converge towards the stable manifold of the NHIM and after a certain time interval we can be sure that the collection of all these trajectories represents quite well the stable manifold of the NHIM.

Whenever one of these trajectories comes close to the $(x,z)$ plane, then the corresponding point on this plane has been marked. The condition for coming close has been chosen as $y = 0$ and $|p_x| < 0.02$ and $|p_y| < 0.02$. I.e. we mark points on the plane which are closer to the stable manifold of the NHIM in momentum direction than these threshold values. In addition, we only include points for which the trajectory has run a time longer than some threshold value $T_s$ in order to be sure that the trajectory is already very close to the stable manifold. The result of the numerical test is the following: All points which are marked yellow or magenta in the CCDs lie in the neighbourhood of points also marked in the intersection plot. Some boundaries of the red and green regions in the CCDs are not marked in the intersection plot because of the following: Red and green in the CCDs means rather fast escape and when the escape time is smaller than the time threshold $T_s$, then also the approach of the stable manifold to this boundary point takes a time less than $T_s$ and according to the rules mentioned above the point is not marked in the intersection plot. This happens in particular for the higher energy -2700 and for boundaries close to the saddle region of the position space. In particular, we note that the interior of the blue, red and green regions in the CCDs is free of points marked in the corresponding intersection plot.

We have used the initial conditions close to one saddle NHIM only. This is sufficient because in the long run the tendrils of the stable manifolds of the other NHIM run towards the same limit set. Or expressed the other way around: The limit sets of
the tendrils of stable manifolds from the NHIMs from both sides coincide. In this sense it is sufficient to look at the stable manifold coming from one of the saddles. In total the comparison of these plots demonstrates graphically the coincidence of basin boundaries with the stable manifolds of the saddle NHIMs.

\section{Scenario of the fundamental vertical orbit}
\label{fvo}

In \citetalias{JZ16b} the development scenario of the dynamics has been described in detail for the parameter values $a = 10$ and $\Omega_{\rm b} = 4.5$. It has be found that the most important invariant subsets in the phase space are the codimension-2 NHIMs over the index-1 saddles of the effective potential. And inside of these NHIMs the most important periodic orbits are the vertical Lyapunov orbits. They direct the whole development scenario to a surprisingly large extent. Therefore we will now describe in detail the effect of changes of the bar parameters $a$ and $\Omega_{\rm b}$ on the development scenario of these most important periodic orbits. Because of symmetry reasons the two NHIMs over the two index-1 saddles (Lagrange points $L_2$ and $L_3$) are equivalent and also their vertical Lyapunov orbits are related by symmetry and thereby equal. Accordingly in the following we simply speak of the codimension-2 NHIM over the index-1 saddle of the effective potential and of its fundamental vertical periodic orbit, which we call $\Gamma_v$ in the following.

Let us start with a brief repetition of the development scenario of $\Gamma_v$ under an increase of the energy as explained in more detail in \citetalias{JZ16b}. When we speak in the following of tangential directions and of normal directions then this always means directions relative to the NHIM surface. It is easiest to study the dynamics of $\Gamma_v$ in particular and of the whole NHIM in general in a Poincar\'{e} map with the intersection condition $z = 0$ where the orientation is irrelevant because of symmetry reasons. $\Gamma_v$ is born at the saddle energy $E_s \approx -3242$ as tangentially stable and normally unstable. Near $E = -3223$ it suffers a first pitchfork bifurcation where it becomes tangentially unstable and splits off a pair of tilted loop orbits. The tangential instability is always small compared to the normal instability, such that $\Gamma_v$ remains normally hyperbolic and stays a part of the NHIM. Near $E = -3214$ it suffers a second pitchfork bifurcation where it returns to tangential stability and splits off a second pair of tilted loop orbits.

Next between $E \approx -3120$ and $E \approx -3108$ the vertical fundamental orbit shows a scenario which at first sight appears complicated and which has not been analysed completely in \citetalias{JZ16b}. First at $E \approx -3120$ two new vertical orbits of period 1 in the Poincar\'{e} map are created in a saddle-centre bifurcation. They lie at smaller $x$ values than $\Gamma_v$. In \citetalias{JZ16b} these two new orbits have been called $\Gamma_c$ and $\Gamma_d$, where $\Gamma_d$ is the inner one. With increasing values of $E$ $\Gamma_d$ moves to still smaller values of $x$ whereas $\Gamma_c$ moves to larger values of $x$ and at $E \approx -3108$ the two orbits $\Gamma_c$ and $\Gamma_v$ collide and destroy each other in a saddle-centre bifurcation. In total the orbit $\Gamma_v$ is replaced by the very similar orbit $\Gamma_d$, which lies further in the inside of the potential hole. The orbit $\Gamma_d$ continues to rather high values of $E$ which are no longer interesting for the escape dynamics. Therefore we do not discuss the behaviour for these high values. Most important, in the energy interval between $E \approx -3120$ and $E \approx -3108$ we find the coexistence of three different fundamental vertical periodic orbits. And in the following we will understand in detail the dependence of this triplication on the bar parameters. The scenario described so far is shown graphically in \citetalias{JZ16b} where Fig. 4 gives the bifurcation diagram, Fig. 5 shows the stability properties and Fig. 6 shows the orbits in position space.

Let us give a comment to Fig. 4 of \citetalias{JZ16b}. It shows as function of $E$ the $x$-coordinates of the fixed points in the Poincar\'{e} map corresponding to the orbits involved in the scenario of $\Gamma_v$ as mentioned above. This $x$ value for the orbit $\Gamma_v$ will frequently be called $x_0$ in the following. If we describe the scenario as function of $E$ then we have the difficulty that in the energy interval between -3120 and -3108 there are three different fundamental vertical orbits. However, we could also interpret this plot as showing the resulting energy value when looking for the fundamental vertical orbit which has a particular $x$ coordinate of its corresponding fixed point in the map. Interpreted this way there is a single value of $E$ for each $x$ value in the interval from 7 to $x_s$, look at the smooth curve formed by the union of the dark green, brown and orange segments in this figure. This joint curve is a graph over the $x_0$ axis. In this sense the orbits $\Gamma_v$, $\Gamma_c$ and $\Gamma_d$ are in reality a single fundamental vertical orbit only which we still call $\Gamma_v$. And only the projection of this curve on the $E$ axis switches between 1:1 for some energy intervals and 3:1 for another energy interval. In the following we will study in detail this singular projection behaviour when we include the dependence of the scenario on the further bar parameters $a$ and $\Omega_{\rm b}$.

\subsection{Dependence on $\Omega_{\rm b}$ for fixed $a = 10$}
\label{ss1}

\begin{figure}
\begin{center}
\includegraphics[width=\hsize]{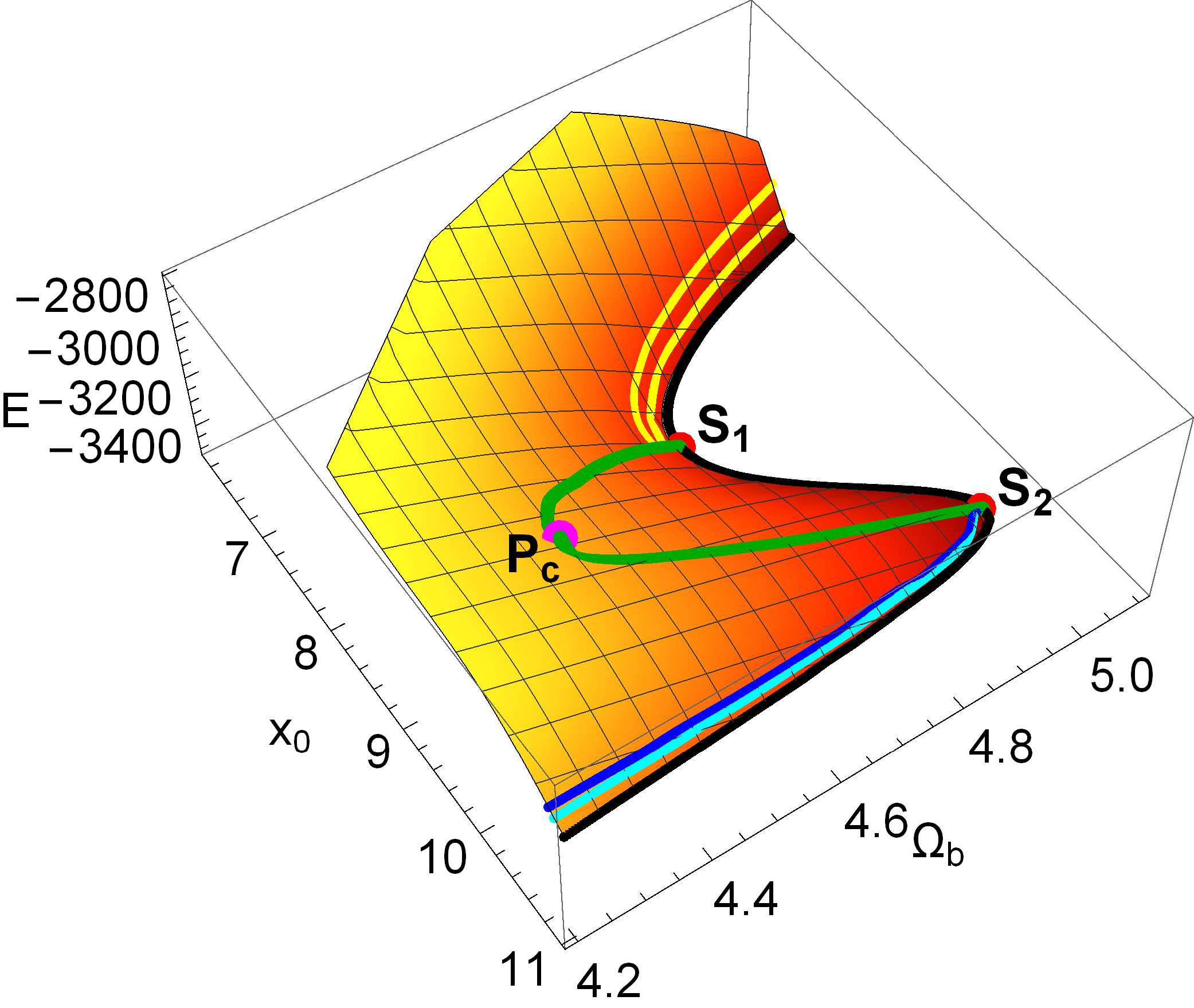}
\end{center}
\caption{Bifurcation diagram of the fundamental vertical orbit $\Gamma_v$. The coordinate $x_0$ is the $x$-coordinate of the fixed point in the Poincar\'{e} map corresponding to the orbit. For each point over the horizontal plane with coordinates $\Omega_{\rm b}$ and $x_0$ the value of the energy $E$ is indicated for which $\Gamma_v$ is realized over this point. The result is a graph over a part of the horizontal plane. The black curve is the boundary of existence of $\Gamma_v$. On this boundary curve the two extremal points in the $\Omega_{\rm b}$ coordinate are called $S_1$ and $S_2$, respectively and they are marked as red points. The cyan, blue and yellow curves mark pitchfork bifurcations of $\Gamma_v$ where it splits off tilted loop orbits. The green curve marks the saddle-centre bifurcations between various branches of $\Gamma_v$. The magenta point, labelled $P_c$, marks the cusp point. The green curve ends in the points $S_1$ and $S_2$ on the black curve. We always use $a = 10$. (For the interpretation of references to colour in this figure caption and the corresponding text, the reader is referred to the electronic version of the article.)}
\label{surf}
\end{figure}

First let us study the dependence of the scenario of $\Gamma_v$ on $\Omega_{\rm b}$ when we still fix $a$ on the value 10. We start by giving in Fig. \ref{surf} a higher dimensional generalization of Fig. 4 from \citetalias{JZ16b}. We start with the 2-dimensional plane of the quantities $x_0$ and $\Omega_{\rm b}$ and plot over this plane the corresponding energy value at which the fundamental vertical orbit is found whenever it exists over this point of the horizontal plane. We obtain a smooth graph over a part of the $(x_0, \Omega_{\rm b})$ plane. We call this surface $S(x_0,\Omega_{\rm b})$. The cut at $\Omega_{\rm b} = 4.5$ reproduces the dark green/brown/orange curve from Fig. 4 of \citetalias{JZ16b}. In the figure we have plotted by colour various important curves on the surface $S(x_0,\Omega_{\rm b})$ and marked a few still more important points. The cyan curve is the line where the first pitchfork bifurcation of $\Gamma_v$ occurs and the blue line indicates its second pitchfork bifurcation. The two yellow curves mark the analogous pitchfork bifurcations on the inner branch of the vertical orbit. The black line is the boundary of the surface, it is the boundary where the existence of $\Gamma_v$ ends because of energetic reasons, i.e. where this orbit ends on an extremal point of the effective potential. Therefore this black curve indicates at the same time the relative extremal points of the effective potential lying along the positive $x$-axis. The green curve marks the saddle-centre bifurcations between the various branches of the vertical orbit when we study the scenario as a function of $E$ as we did in \citetalias{JZ16b}. This green curve ends on the points of the boundary curve which are extremal points in $\Omega_{\rm b}$ direction. We call these two points $S_1$ and $S_2$ and we mark them by red dots. The coordinates of these two end points are $x_{0,s1} = 8.28$, $\Omega_{b,s1} = 4.71$, $E_{s1} = -3381$ and $x_{0,s2} = 9.77$, $\Omega_{b,s2} = 5.02$, $E_{s2}= - 3507$, respectively. The green curve itself also has an extremal point in $\Omega_{\rm b}$ direction, it is marked as a magenta dot and we call it $P_c$. Its coordinates are $x_{0,c} = 9.1$, $\Omega_{b,c} = 4.42$ and $E_c = -3029$. As we will see in a moment the magenta point is a cusp point of the scenario where in $\Omega_{\rm b}$ direction the existence of the region with three branches of the vertical orbit ends.

The part of the surface of Fig. \ref{surf} which lies inside of the green bifurcation curve describes orbits which are normally elliptic and which therefore can not belong to the NHIM. The rest of the surface represents orbits which are normally hyperbolic and which belong to the NHIM as long as the normal instability is larger than any possible tangential instability. We find such weak tangential instability in the regions between the two pitchfork curves. Please note that according to the previous paragraph there are three different extremal points of the effective potential along the positive $x$-axis in the $\Omega_{\rm b}$ interval $(\Omega_{b,s1},\Omega_{b,s2})$. The outermost and the innermost one are index-1 saddles and the middle one is a relative minimum of the potential, formally we can say that the middle one is an index-0 saddle. At $\Omega_{b,s1}$ the inner saddle and the minimum collide and disappear, whereas at $\Omega_{b,s2}$ the outer saddle and the minimum collide and disappear. Thereby in the end with increasing values of $\Omega_{\rm b}$ the saddle point lying at large values of $x$ is replaced by a new saddle point lying further inside at smaller values of $x$.

\begin{figure}
\begin{center}
\includegraphics[width=\hsize]{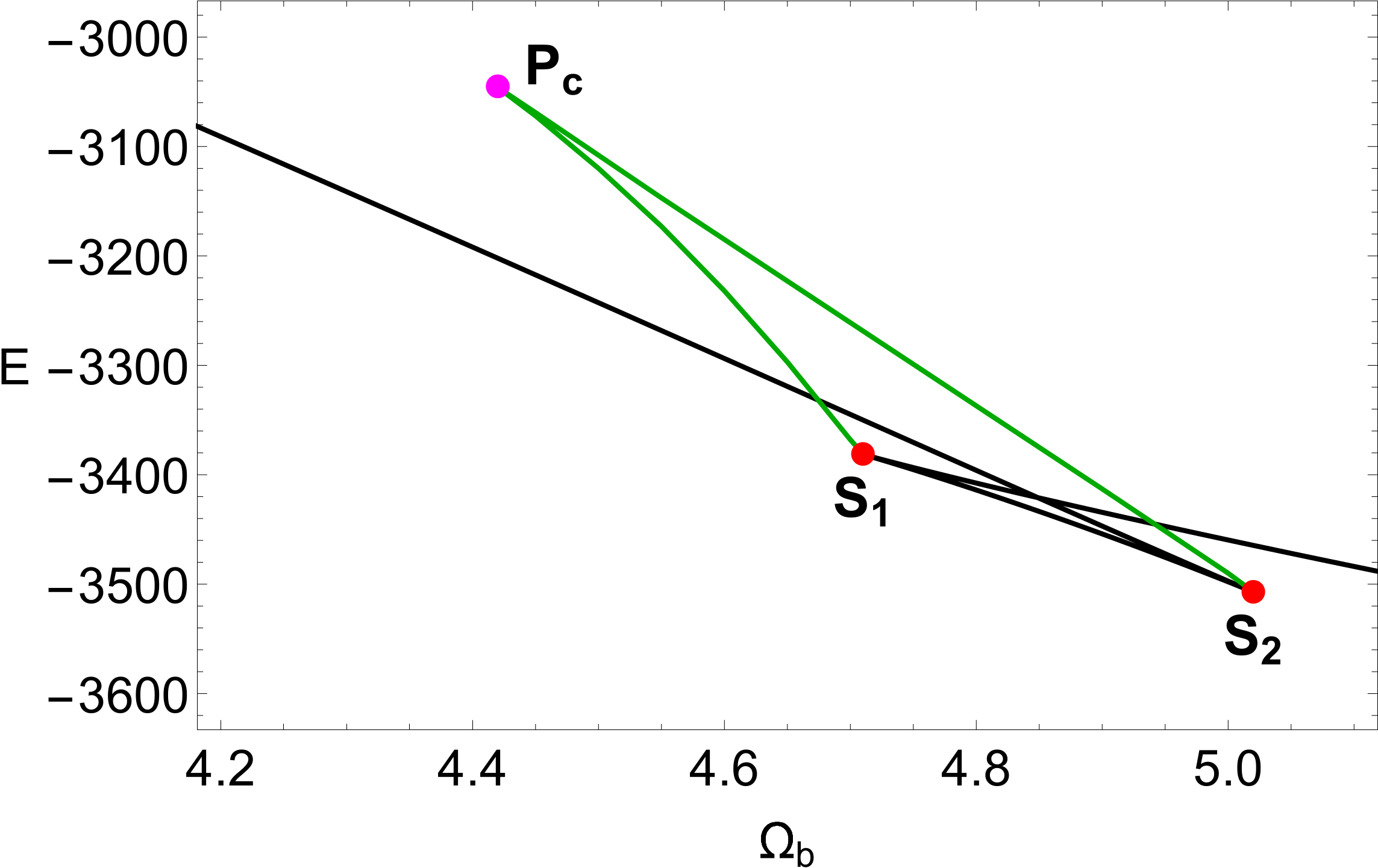}
\end{center}
\caption{Projection of the black and green curves from Fig. \ref{surf} into the $(\Omega_{\rm b},E)$ plane, when $a = 10$. The magenta point $P_c$ and the red points $S_1$ and $S_2$ are the projections of the equally labelled points from Fig. \ref{surf}. (For the interpretation of references to colour in this figure caption and the corresponding text, the reader is referred to the electronic version of the article.)}
\label{cusp}
\end{figure}

By definition the black boundary curve and the green saddle-centre bifurcation curve on the surface $S(x_0,\Omega_{\rm b})$ indicate the parameter values where the various branches of the fundamental vertical orbit are created and destroyed. To see this bifurcation behaviour clearer we project now the surface $S(x_0,\Omega_{\rm b})$ into the $(\Omega_{\rm b},E)$ plane and in particular we keep track of the black boundary curve and of the green saddle-centre bifurcation curve. The result is given in Fig. \ref{cusp}. Corresponding points are marked by dots of the same colour and labelled equal in the two figures \ref{surf} and \ref{cusp}. Now it becomes evident at first sight in which sense the point $P_c$ becomes a cusp point. The various curve segments in this plot of Fig. \ref{cusp} tell us how the various branches of the fundamental vertical orbit come and go under parameter changes. The segment of the black curve coming from the left and ending in the point $S_2$ indicates where the outer hyperbolic vertical orbit ends on the outer index-1 saddle point of the effective potential. The segment of the black curve coming from the right and ending in the point $S_1$ indicates where the inner hyperbolic vertical orbit ends on the inner index-1 saddle point of the potential. The middle segment of the black curve indicates where the normally elliptic vertical orbit ends in the minimum of the potential. The lower segment of the green curve indicates along which parameter curve the inner hyperbolic and the middle elliptic vertical orbits collide and destroy each other in a saddle-centre bifurcation. The upper segment of the green curve indicates for which parameter values the outer hyperbolic and the middle elliptic vertical orbit collide and destroy each other in a saddle-centre bifurcation. For the cusp point of the green curve all three vertical orbits collide and transform each other into a single hyperbolic vertical orbit. This is an exceptional bifurcation and therefore the point $P_c$ is a singular point of the scenario. However in the Figs. \ref{surf} and \ref{cusp} this singularity is unfolded by the embedding of this point into a higher-dimensional parameter space. More general information on cusp singularities can be found in section 9.5 of \citet{PS78}.

In Fig. \ref{cusp} it is demonstrated in a very clear form how both segments of the green saddle-centre curve end in the cusp points $S_1$ and $S_2$ appearing in the projection of the black boundary curve. I.e. in each one of these two points meet two segments of the boundary curve and one segment of the saddle-centre bifurcation curve. The relation between the two cusps of
the boundary curve and the cusp of the saddle-centre curve will become evident in the next subsection when we go to a still higher dimensional parameter space.

Also note the following: For each fixed value of $\Omega_{\rm b}$ between $\Omega_{b,c}$ and $\Omega_{b,s2}$ there are two disjoint versions of the normally hyperbolic vertical orbits, the ones which we have called $\Gamma_v$ and $\Gamma_d$ in \citetalias{JZ16b}. For fixed $\Omega_{\rm b} \in (\Omega_{b,c},\Omega_{b,s1})$ we can only connect these two versions by running through the interior of the green saddle-centre bifurcation curve and therefore running through two saddle-centre bifurcations involving the orbit $\Gamma_c$. However in the scenario which includes the variation of the parameter $\Omega_{\rm b}$ we can connect the two normally hyperbolic versions of the vertical orbit along a path in the parameter space which encircles the green curve and thereby avoids the interior of the green curve and avoids running through the normally elliptic region of the vertical orbit. So we connect the two hyperbolic branches of the vertical orbit smoothly without ever running through any saddle-centre bifurcation of the various branches of the vertical orbit. In this sense, the orbits $\Gamma_v$ and $\Gamma_d$, appearing in \citetalias{JZ16b}, are in reality one and the same fundamental vertical periodic orbit.

Finally we have to make a comment on the region of $\Omega_{\rm b}$ values where our model makes sense. The Lagrange radius should be larger than the bar's semi-major axis $a$. As an estimate of the Lagrange radius we take the $x$ coordinate of the outer saddle point of the effective potential, i.e. $L_2$. For $a = 10$ its dependence on $\Omega_{\rm b}$ is represented by the outer segment of the black curve in Fig. \ref{surf}. The $x$ coordinate of this curve is larger than 10 for $\Omega_{\rm b} < 4.98$. In this sense, we can claim that for $a = 10 $ our model makes sense for $\Omega_{\rm b} $ values up to approximately 4.98. This includes a major part of the green bifurcation curve of Figs. \ref{surf} and \ref{cusp} and in particular it includes the cusp point with all the complications caused by it. Also the whole scenario described in \citetalias{JZ16b} is included in the region where the model is realistic.

The complicated bifurcation scenario of the fundamental vertical orbit comes to a large part from the complicated scenario of the Lagrange point $L_2$ and its possibility to triplicate in some $\Omega_{\rm b}$ interval. This triplication of the Lagrange point $L_2$ in a barred galaxy model has also been observed in \citet{ARGM09} (see e.g., Fig. 12). However in this publication the triplication has been forced by an addition to the effective potential. In our model on the other hand, this scenario is automatically included in the basic structure of the bar potential. Loosely speaking for increasing values of $\Omega_{\rm b}$ first the triplication of the saddle point and finally the replacement of the outer saddle by another one lying further inside occurs when the increasing centrifugal potential dominates the bar potential also inside of the bar itself.

\subsection{Simultaneous dependence on $\Omega_{\rm b}$ and $a$}
\label{ss2}

\begin{figure}
\begin{center}
\includegraphics[width=\hsize]{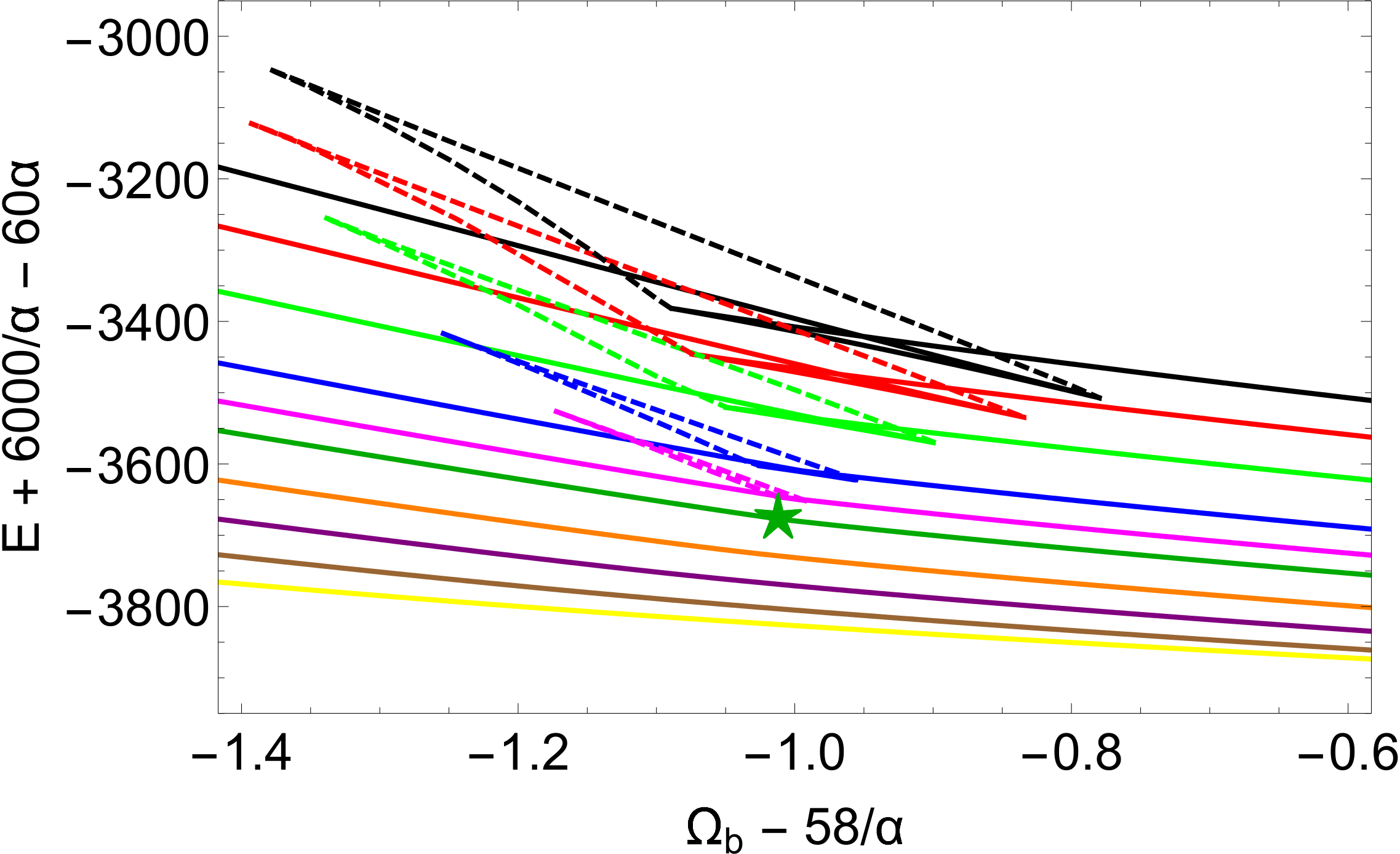}
\end{center}
\caption{The same construction as in Fig. \ref{cusp}, with the curves for various values of $a$ included. The solid lines are the projections of the boundary curves and the broken lines are the projections of the saddle-centre bifurcation curves. The $a$ values are colour coded as: $a = 10$ (black); $a = 9$ (red); $a = 8$ (green); $a = 7$ (blue); $a = 6.5$ (magenta); $a = 6.123$ (dark green); $a = 5.5$ (orange); $a = 5$ (purple); $a = 4.5$ (brown); $a = 4$ (yellow). The swallowtail point is marked by a 5-pointed dark green star. The curves are plotted in transformed coordinates (for more details regarding the
transformation see the axis labels and the main text). (For the interpretation of references to colour in this figure caption and the corresponding text, the reader is referred to the electronic version of the article.)}
\label{cusp2}
\end{figure}

Finally we study how the scenario depends on the bar length $a$. A plot of the energy surface created by the fundamental vertical periodic orbit over the 3-dimensional $(x_0,\Omega_{\rm b},a)$ space would be a graph in a 4-dimensional total space and could not be represented in any understandable form as a 2-dimensional plot.

Therefore we start by a higher dimensional analogue of Fig. \ref{cusp} where the parameter $a$ is included by colour. If we would plot all the curves for the various values of $a$ into the plane with the coordinates $\Omega_{\rm b}$ and $E$ then the interesting and important parts of these curves would be distributed over a large area (large compared to the relatively small size of the interesting part of each individual curve) and the plot would be difficult to understand. Therefore, we introduce an $a$ dependent transformation of the coordinates $\Omega_{\rm b}$ and $E$ in order to shift the important parts of the individual curves and to bring them into a relative position to each other which makes it easy to understand the global structure. This transformation is $\Omega_{\rm b} \to \Omega_{\rm b} - 58 / a$ and $E \to E + 6000/a - 60 a$. The numerical values in this transformation do not have any deeper meaning. They are just chosen to produce a convenient presentation of the plot. The result of plotting in these transformed coordinates is presented in Fig. \ref{cusp2}. The relation between colour and values of $a$ is as follows: Black corresponds to $a = 10$, red to $a = 9$, green to $a = 8$, blue to $a = 7$, magenta to $a = 6.5$, dark green to $a = 6.123$, orange to $a = 5.5$, purple to $a = 5$, brown to $a = 4.5$ and yellow to $a = 4$. The projection of the boundary curve into the transformed $(\Omega_{\rm b},E)$ plane is given as solid curves and the projection of the bifurcation curve into this plane is given as broken curve.

Instead of the 10 values used for $a$ only we should imagine in our mind the whole continuum of $a$ values and the corresponding interpolated solid and broken curves. Thereby we obtain several 2-dimensional surfaces embedded into the 3-dimensional parameter space. The surfaces formed by the continuum of the solid curves are a swallowtail surface well known from catastrophe theory (see section 9.5 of the book by \citet{PS78} and the figures in it). The cusp lines of this surface meet and end in the swallowtail point at $a_{s} \approx 6.123$ which is marked by an 5-pointed star in Fig. \ref{cusp2}. This point is at the same time also the end point of the cusp line coming from the cusps of the bifurcation surface built up by the continuum of the broken lines. Note that the boundary lines of the bifurcation surface coincide with the cusp lines of the boundary surface.

\begin{figure}
\begin{center}
\includegraphics[width=\hsize]{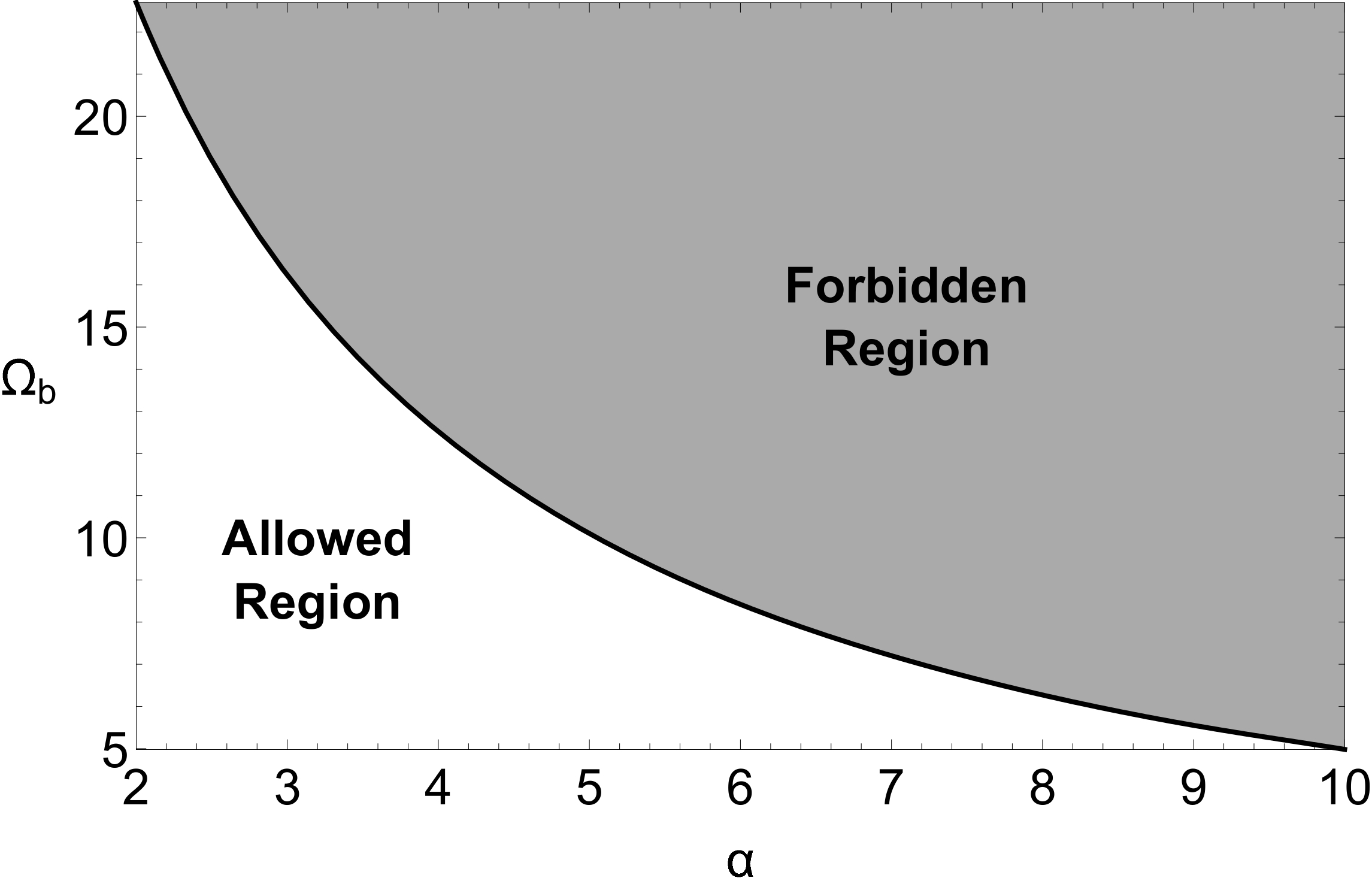}
\end{center}
\caption{Division of the $(a, \Omega_{\rm b})$ parameter plane into the parts where our model makes sense (white areas) and where it does not (grey-shaded areas). The criterion for the allowed region is that the Lagrange radius $x_s$, i.e. the
$x$-coordinate of the outer index-1 saddle of the effective potential, is larger than the bar length $a$. (For the interpretation of references to colour in this figure caption and the corresponding text, the reader is referred to the electronic version of the article.)}
\label{real}
\end{figure}

\begin{figure*}
\centering
\resizebox{\hsize}{!}{\includegraphics{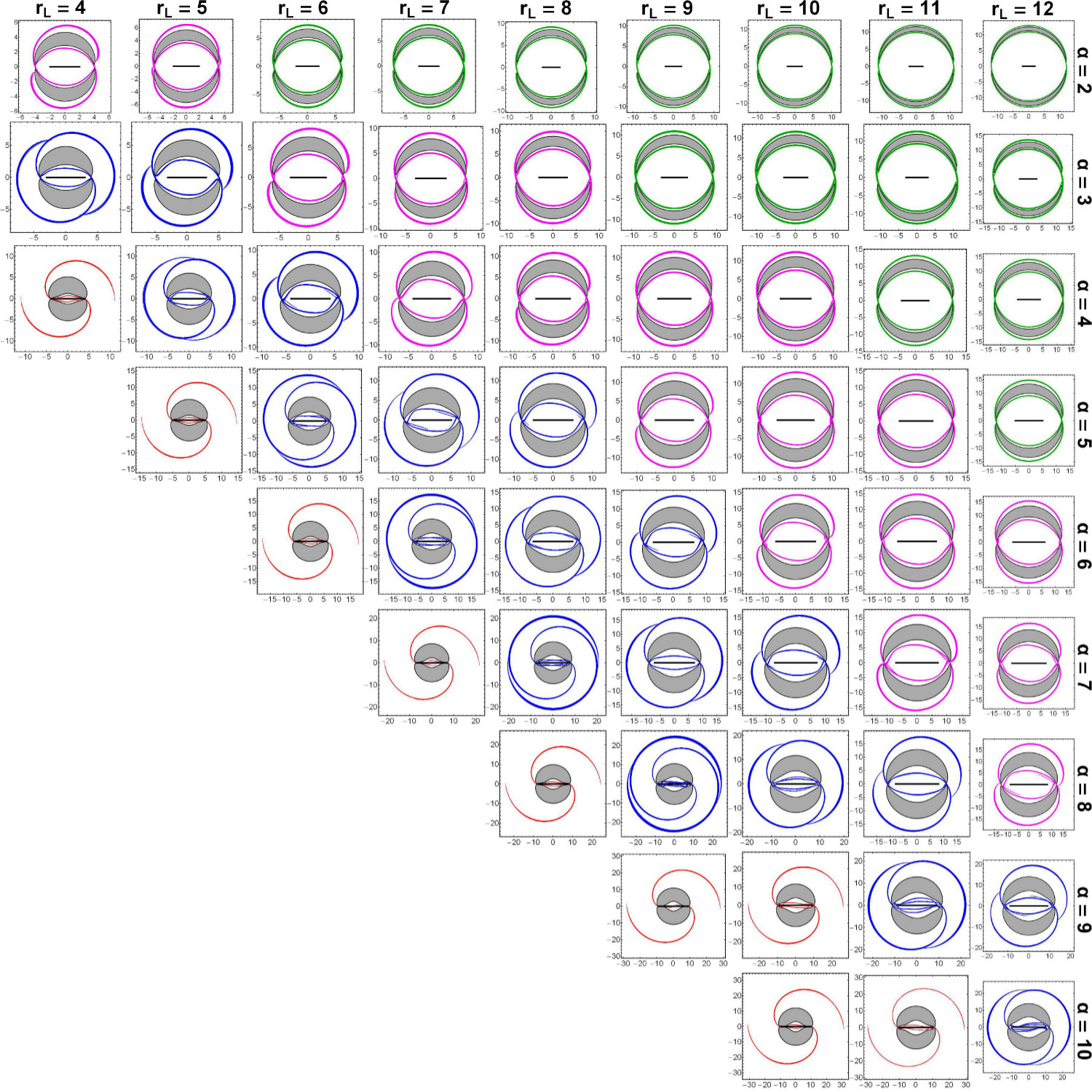}}
\caption{A collection of morphologies of the unstable manifolds for several values of the semi-major axis $a$ of the galactic bar as well as of the Lagrange radius $r_L$. For all models we have $\widehat{C} = 0.0001$. The manifolds are plotted in a different colour which is determined by the corresponding morphology. The colour code is as follows: $R_1$ rings (green); $R_1'$ pseudo-rings (magenta); $R_1R_2$ rings (blue); open spirals (red). The horizontal black lines in the interior region indicate the total length of the bar. (For the interpretation of references to colour in this figure caption and the corresponding text, the reader is referred to the electronic version of the article.)}
\label{mans}
\end{figure*}

When we make the same estimates which we made at the end of the previous subsection for the part of the parameter space in which our model makes sense then the result is presented in Fig. \ref{real}. It shows a boundary line which divides the 2-dimensional $(a,\Omega_{\rm b})$ plane into a forbidden part at large values of these parameters and an allowed part for small values. Forbidden values of the parameters simply mean that these values are not realistic for a barred galaxy model.

As essential result we note the following: If the bar is larger than the swallowtail value $a_s$, then there is an $\Omega_{\rm b}$ interval in which the effective potential has 3 different Lagrange points along the positive $x$-axis, of course because of symmetry reasons there are also the symmetrical ones along the negative $x$-axis. This leads to a corresponding triplication of the fundamental vertical periodic orbit in a certain energy interval as described in more detail in \citetalias{JZ16b}, for the special case of $a = 10$ and $\Omega_{\rm b} = 4.5$.

\section{Formation of rings and spirals}
\label{spr}

In Papers I and II (see Fig. 14 in \citetalias{JZ16a} and Fig. 11 in \citetalias{JZ16b}) we proved that in both the 2-dof and 3-dof systems, respectively, the value of the semi-major axis of the bar $a$ strongly influences the structure of the unstable manifolds, which determines the observed stellar structure formed by stars escaping over the saddles $L_2$ and $L_3$.

Now, an interesting question that rises is: How does the shape of the stellar structures, guided by the outer manifolds, depend on the parameters determining the bar potential? To answer this question we performed numerous tests by varying the values of the all the parameters related with the barred structure (that is the mass and the scale length of the bar and of course the angular velocity of the bar). Our calculations strongly suggest that the most influential parameter, after the semi-major axis of the bar, is its angular velocity, while the mass as well as the scale length play secondary roles regarding the shapes of the manifolds.

In Fig. \ref{mans} we present the local segments of the unstable manifolds around $L_2$ and $L_3$ for a large variety of galaxy models, where the pattern speed $\Omega_{\rm b}$ is parameterized by the value of the Lagrange radius $r_L$ \citep[see also][]{ARGBM09,ARGBM10,ARGM11,RGAM07,RGMA06}. For every values of $a$ and $\Omega_{\rm b}$ the energy level $E$ is chosen such that it is $\widehat{C} = 0.0001$\footnote{The energy of escape $E(L_2)$ can be used in order to define a dimensionless energy parameter as $\widehat{C} = \left(E(L_2) - E\right)/E(L_2)$, where $E$ is some other value of the energy integral. This dimensionless energy parameter $\widehat{C}$ makes more convenient the reference to energy levels above the escape energy.} above the respective saddle energy which also depends on the values of the parameters of the bar. Note, that according to Fig. \ref{real}, for a galaxy model to be realistic it should be $a \leq r_L$.

For constructing the plots of the morphologies shown in the multi-panel Fig. \ref{mans} we worked as follows: first we define a set of 5000 equally spaced initial conditions inside the 3-dimensional NHIM hyper-surface, near the saddle point $L_2$. Then we numerically integrate them forward in time and we let them align with the unstable manifold, while keeping records for all the coordinates and the momenta of the orbits. The numerical integration stops only when the full pattern of the respective morphology is developed. For symmetry reasons we can automatically obtain the data regarding the manifolds around $L_3$, without additional computations. In Fig. \ref{mans} we decided to plot the projection of the manifolds on the galactic $(x,y)$ plane only for clearer representation of the stellar morphologies. Our analysis suggests that these structures live very close the galactic plane, since the average value of the $z$ coordinate of the orbits was found equal to 0.05. Additional simulations for values of energy larger than $\widehat{C} = 0.0001$ indicate that the overall types of the stellar structures remain the same. The only observable difference is the increase of the average value of the $z$ coordinate of the orbits.

At this point, we would like to stress that the existence of the unstable manifolds around the index-1 saddle points is only a necessary but by no means a sufficient condition for the corresponding stellar structure to develop. This is true if we think of the following argument: In theory, a manifold may be present (obtained by the numerical integration) for a particular galaxy model. However, in a real barred galaxy, with similar dynamical properties (like those taken into account in the corresponding mathematical model) the manifold may not be able to trap inside of it a sufficient amount of escaping stars and therefore the corresponding stellar structure will not be observable.

Looking at the development scenario presented in Fig. \ref{mans} it becomes evident that the resulting stellar structure clearly depends on both the semi-major axis of the bar and the pattern speed. In particular, we can define a quantity, called stellar formation index, as SFI = $r_L/a$. Then we may conclude that:
\begin{itemize}
  \item When SFI $> 2.6$ the local segments of the unstable manifolds trace out a ring structure around the interior region of the galaxy. The structure where the major axis of the ring points into the $y$ direction is called $R_1$ ring. In this case, the unstable manifolds from one side come very close to the opposite saddle point thus forming approximate heteroclinic separatrix connections.
  \item When SFI $\in [1.5,2.6)$ the heteroclinic connections are clearly broken. Now the major axis of the ring rotates in negative orientation with increasing value of the semi-major axis of the bar. These structures are called $R_1'$ pseudo-rings.
  \item When SFI $\in (1.1,1.5)$ the unstable manifolds coming from one side connect to the unstable manifold from the other side in a point far away from the saddle. These structures are called $R_1R_2$ ring. It is seen that the major axis of the $R_1R_2$ ring still rotates in negative orientation with increasing value of $a$.
  \item When SFI $\in [1,1.1]$ the orientation of the major axis of the $R_1R_2$ rings approaches the $x$ axis, then the rings break and the unstable manifolds form twin open spirals which begin very close to the two ends of the bar.
\end{itemize}

It should be emphasized that the classification of the morphologies shown in Fig. \ref{mans} was performed by eye inspection, thus following the method according to which observers classify real galaxies. Generally speaking, differentiating between the different types of the stellar morphologies is a rather easy task. However, there are some borderline cases, such as those between $R_1$ and $R_1'$ or between $R_1'$ and $R_1R_2$ for which the classification is a matter of personal judgment. The reader can find more useful information regarding classification of stellar structures in \citet{BC96}.

Taking into account all the results (from all the papers of the series) regarding the stellar formations, we may argue that in our model $R_2$ rings, where the major axis points into the bar direction, cannot be developed by any combination of the parameters of the bar.

\section{Discussion}
\label{disc}

In this paper we numerically investigated the escape dynamics of a 3-dof dynamical model which describes the motion of stars in a barred galaxy with a spherically symmetric central nucleus, a flat thin disc and dark matter halo component. A thorough and systematic orbit classification of large sets of initial conditions of orbits is conducted in several types of two-dimensional planes in an attempt to obtain a complete view of the orbital dynamics of the system. We managed to reveal the basins of escape and also to relate them with the corresponding escape times of the orbits. Inside of the 5-dimensional energy shells we found well-defined basins of escape which we displayed graphically by their intersections with appropriate 2-dimensional surfaces. The basins have highly fractal boundaries. Regions of bounded regular motion are also detected. Furthermore, we illustrated that the fractal basin boundaries are in reality the intersections of the stable manifolds, around the index-1 saddle points, with the $(x,z)$ plane.

We proved that the 3-dof model of the barred galaxy contains all the important types of orbits which should be present in a valid barred galaxy model. Such orbits are the basic loop orbits (parallel or tilted with respect to the galactic $(x,y)$ plane) and of course the x1 elongated orbits which are the building blocks of the galactic bar.

In \citetalias{JZ16b} we showed that the most important subsets in the phase space are the NHIMs over the index-1 saddle points $L_2$ and $L_3$ of the effective potential. Moreover, inside the NHIMs the vertical Lyapunov periodic orbits is the type of orbits with paramount importance. For this reason, we described in detail how the parameters of the bar, such as the major-semi axis and the angular velocity, influence the development scenario of these orbits, thus revealing the unfolding of the cusp singularity.

Both the stable and the unstable manifolds of the NHIMs direct the flow over the index-1 saddle points and thereby are mainly responsible, to a large extent, for the global stellar structure formations, such as rings and spirals. To demonstrate this we examined how the combination of the most influential parameters of the bar affect the observed stellar formations. Going one step further, we defined a dynamical quantity, called "stellar formation index" (SFI), which allows us to predict the corresponding stellar structure.

The present paper is the third and last paper of the series devoted on the orbital and escape dynamics in barred galaxies as well as on the role of the NHIMs. We hope that the combined results of this series are useful in the field of escaping stars in barred galaxies.

\section*{Acknowledgments}

One of the authors (CJ) thanks DGAPA for financial support under grant number IG-100616. The authors would like to thank the anonymous referee for all the apt suggestions and comments which improved both the quality and the clarity of the paper.

\bsp
\label{lastpage}

\end{document}